\documentclass[twocolumn]{aastex631}

\newcommand{\civd}{C~IV $\lambda\lambda 1548$, 1551}
\newcommand{\oiiit}{O~III] $\lambda\lambda 1661$, 1666}
\newcommand{\nivd}{N~IV] $\lambda\lambda 1483$, 1488}
\newcommand{\oiid}{[O~II] $\lambda\lambda 3726$, 3729}

\newcommand{\ciii}{C~III] $\lambda 1908$}
\newcommand{\civ}{C~IV $\lambda 1549$}
\newcommand{\heii}{He~II $\lambda 1640$}
\newcommand{\oiii}{O~III] $\lambda 1663$}
\newcommand{\oii}{[O~II] $\lambda 3727$}
\newcommand{\niii}{N~III] $\lambda 1750$}
\newcommand{\niv}{N~IV] $\lambda 1488$}
\newcommand{\neiii}{[Ne~III] $\lambda 3868$}
\newcommand{\neiv}{[Ne~IV] $\lambda 2424$}
\newcommand{\nev}{[Ne~V] $\lambda 3426$}
\newcommand{\oiiibowen}{O~III $\lambda 3133$}
\newcommand{\logoh}{12 + log(O/H)}
\newcommand{\zzsun}{$Z/{\rm Z}_{\odot}$}
\newcommand{\ergs}{erg~s$^{-1}$~cm$^{-2}$}
\newcommand{\ergsA}{erg~s$^{-1}$~cm$^{-2}$ \AA$^{-1}$}

\newcommand{\pyneb}{\texttt{PyNeb}}

\newcommand{\rxc}{RXCJ2248-ID}
\newcommand{\AStark}{A1703-zd6}
\newcommand{\JIzotov}{J2229+2727}

\shorttitle{Spectroscopic confirmation of GHZ2 at z=12.34}
\shortauthors{Castellano et al.}
\usepackage[autopunct=true]{csquotes}

\usepackage{amsmath}
\usepackage{graphicx}
\usepackage{natbib}
\usepackage{scalerel}
\usepackage{makecell}
\usepackage{multirow}
\usepackage{txfonts}

\usepackage{hyperref}  
 \usepackage{color}
\definecolor{blue}{rgb}{0., 0., 1}
\usepackage{rotating,tabularx}
\hypersetup{colorlinks=true,linkcolor=[rgb]{1.,0.2,0.2},citecolor=[rgb]{0.1,0.4,1.},filecolor=[rgb]{0.7,0.2,0.2},urlcolor=[rgb]{0.7,0.2,0.2}}

\begin{document}
\title{JWST NIRSpec Spectroscopy of the Remarkable Bright Galaxy GHZ2/GLASS-z12 at Redshift 12.34}

\correspondingauthor{Marco Castellano}
\email{marco.castellano@inaf.it}

\author[0000-0001-9875-8263]{Marco Castellano}
\affiliation{INAF - Osservatorio Astronomico di Roma, via di Frascati 33, 00078 Monte Porzio Catone, Italy}

 \author[0000-0002-8951-4408]{Lorenzo Napolitano}
 \affiliation{INAF - Osservatorio Astronomico di Roma, via di Frascati 33, 00078 Monte Porzio Catone, Italy}
 \affiliation{Dipartimento di Fisica, Università di Roma Sapienza, Città Universitaria di Roma - Sapienza, Piazzale Aldo Moro, 2, 00185, Roma, Italy}

 \author[0000-0003-3820-2823]{Adriano Fontana}
 \affiliation{INAF - Osservatorio Astronomico di Roma, via di Frascati 33, 00078 Monte Porzio Catone, Italy}

\author[0000-0002-4140-1367]{Guido Roberts-Borsani}
\affiliation{Department of Astronomy, University of Geneva, Chemin Pegasi 51, 1290 Versoix, Switzerland}

 \author[0000-0002-8460-0390]{Tommaso Treu}
 \affiliation{Department of Physics and Astronomy, University of California, Los Angeles, 430 Portola Plaza, Los Angeles, CA 90095, USA}

\author[0000-0002-5057-135X]{Eros Vanzella}
\affiliation{INAF -- OAS, Osservatorio di Astrofisica e Scienza dello Spazio di Bologna, via Gobetti 93/3, I-40129 Bologna, Italy}

\author[0000-0002-7051-1100]{Jorge A. Zavala}
\affil{National Astronomical Observatory of Japan, 2-21-1, Osawa, Mitaka, Tokyo, Japan}

\author[0000-0002-7959-8783]{Pablo Arrabal Haro}
\affil{NSF’s National Optical-Infrared Astronomy Research Laboratory, 950 N. Cherry Ave., Tucson, AZ 85719, USA;}

 \author[0000-0003-2536-1614]{Antonello Calabr\`o}
 \affiliation{INAF - Osservatorio Astronomico di Roma, via di Frascati 33, 00078 Monte Porzio Catone, Italy}

\author[0000-0003-1354-4296]{Mario Llerena}
\affiliation{INAF - Osservatorio Astronomico di Roma, via di Frascati 33, 00078 Monte Porzio Catone, Italy}

\author[0000-0002-9572-7813]{Sara Mascia}
\affiliation{INAF - Osservatorio Astronomico di Roma, via di Frascati 33, 00078 Monte Porzio Catone, Italy}

\author[0000-0001-6870-8900]{Emiliano Merlin}
\affiliation{INAF - Osservatorio Astronomico di Roma, via di Frascati 33, 00078 Monte Porzio Catone, Italy}

\author[0000-0002-7409-8114]{Diego Paris}
 \affiliation{INAF - Osservatorio Astronomico di Roma, via di Frascati 33, 00078 Monte Porzio Catone, Italy}

\author[0000-0001-8940-6768 ]{Laura Pentericci}
\affiliation{INAF - Osservatorio Astronomico di Roma, via di Frascati 33, 00078 Monte Porzio Catone, Italy}

 \author[0000-0002-9334-8705]{Paola Santini}
 \affiliation{INAF - Osservatorio Astronomico di Roma, via di Frascati 33, 00078 Monte Porzio Catone, Italy}

\author[0000-0002-5268-2221]{Tom J.L.C. Bakx} 
\affil{Department of Space, Earth, \& Environment, Chalmers University of Technology, Chalmersplatsen 4 412 96 Gothenburg, Sweden}

\author[0000-0003-1383-9414]{Pietro Bergamini}
\affiliation{Dipartimento di Fisica, Università degli Studi di Milano, via Celoria 16, I-20133 Milano, Italy}
\affiliation{INAF - OAS, Osservatorio di Astrofisica e Scienza dello Spazio di Bologna, via Gobetti 93/3, I-40129 Bologna, Italy}

\author[0000-0002-6830-9093]{Guido Cupani}
\affil{INAF - Osservatorio Astronomico di Trieste, via Tiepolo 11, I-34143 Trieste, Italy}
\affil{IFPU - Institute for Fundamental Physics of the Universe, via Beirut 2, I-34151 Trieste, Italy}

\author[0000-0001-5414-5131]{Mark Dickinson}
\affiliation{NSF’s NOIRLab, Tucson, AZ 85719, USA}

\author[0000-0003-3460-0103]{Alexei V. Filippenko}
\affiliation{Department of Astronomy, University of California, Berkeley, CA 94720-3411, USA}

\author[0000-0002-3254-9044]{Karl Glazebrook}
\affiliation{Centre for Astrophysics and Supercomputing, Swinburne University of Technology, PO Box 218, Hawthorn, VIC 3122, Australia}

\author[0000-0002-5926-7143]{Claudio Grillo}
\affiliation{Dipartimento di Fisica, Università degli Studi di Milano, via Celoria 16, I-20133 Milano, Italy}
\affiliation{INAF—IASF Milano, via A. Corti 12, I-20133 Milano, Italy}

\author[0000-0003-3142-997X]{Patrick L. Kelly}
\affiliation{Minnesota Institute for Astrophysics, University of Minnesota, 116 Church St. SE, Minneapolis, MN 55455 USA}

\author[0000-0001-6919-1237]{Matthew A. Malkan}
\affiliation{Department of Physics and Astronomy, University of California, Los Angeles, 430 Portola Plaza, Los Angeles, CA 90095, USA}

\author[0000-0002-3407-1785]{Charlotte A. Mason}
\affiliation{Cosmic Dawn Center (DAWN)}
\affiliation{Niels Bohr Institute, University of Copenhagen, Jagtvej 128, 2200 København N, Denmark}

\author[0000-0002-8512-1404]{Takahiro Morishita}
\affiliation{IPAC, California Institute of Technology, MC 314-6, 1200 E. California Boulevard, Pasadena, CA 91125, USA}

\author[0000-0003-2804-0648 ]{Themiya Nanayakkara}
\affiliation{Centre for Astrophysics and Supercomputing, Swinburne University of Technology, PO Box 218, Hawthorn, VIC 3122, Australia}

\author[0000-0002-6813-0632]{Piero Rosati}
\affiliation{Dipartimento di Fisica e Scienze della Terra, Universit\`a degli Studi di Ferrara, Via Saragat 1, I-44122 Ferrara, Italy}
\affiliation{INAF -- OAS, Osservatorio di Astrofisica e Scienza dello Spazio di Bologna, via Gobetti 93/3, I-40129 Bologna, Italy}

\author[0000-0002-3140-4070]{Eleonora Sani}
\affil{European Southern Observatory, Alonso de Córdova 3107, Vitacura, Región Metropolitana, Chile}

\author[0000-0002-9373-3865]{Xin Wang}
\affil{School of Astronomy and Space Science, University of Chinese Academy of Sciences (UCAS), Beijing 100049, China}
\affil{National Astronomical Observatories, Chinese Academy of Sciences, Beijing 100101, China}
\affil{Institute for Frontiers in Astronomy and Astrophysics, Beijing Normal University,  Beijing 102206, China}

\author[0000-0001-9163-0064]{Ilsang Yoon}
\affil{National Radio Astronomy Observatory, 520 Edgemont Road, Charlottesville, VA 22903, USA}





\begin{abstract}
We spectroscopically confirm the $M_{\rm UV} = -20.5$~mag galaxy GHZ2/GLASS-z12 to be at redshift $z=12.34$. The source was selected via NIRCam photometry in GLASS-JWST ERS data, providing the first evidence of a surprising abundance of bright galaxies at $z \gtrsim 10$.  
The NIRSpec PRISM spectrum shows detections of N~IV, C~IV, He~II, O~III, C~III, O~II, and Ne~III lines, and the first detection at high-redshift of the O~III Bowen fluorescence line at 3133~\AA\ rest-frame. The prominent C~IV line with rest-frame equivalent width (EW) $\approx 46$~\AA\ puts GHZ2 in the category of extreme C~IV emitters. GHZ2 displays UV lines with EWs that are only found in active galactic nuclei (AGNs) or composite objects at low/intermediate redshifts. The UV line-intensity ratios are compatible both with AGNs and star formation in a low-metallicity environment, with the low limit on the [Ne~IV]/[N~IV] ratio favoring a stellar origin of the ionizing photons. We discuss a possible scenario in which the high ionizing output is due to low metallicity stars forming in a  dense environment. We estimate a metallicity $\lesssim 0.1$~\zzsun, a high ionization parameter log\,$U > -2$, a N/O abundance 4--5 times the solar value, and a subsolar C/O ratio similar to the recently discovered class of nitrogen-enhanced objects. Considering its abundance patterns and the high stellar mass density ($10^4$~M$_{\odot}$~pc$^{-2}$), GHZ2 is an ideal formation site for the progenitors of today’s globular clusters. The remarkable brightness of GHZ2 makes it a ``Rosetta stone'' for understanding the physics of galaxy formation within just 360~Myr after the Big Bang. 
\end{abstract}




\section{Introduction}\label{sec:intro}
The search for and characterization of galaxies beyond redshift $z = 9$--10 has been one of the drivers for the development of the {\it James Webb Space Telescope (JWST)}, and the earliest results have not been short of surprises. As demonstrated by a number of different surveys,  \citep[e.g.,][]{Castellano2022b,Castellano2023a,Finkelstein2022b, Finkelstein2023a,Finkelstein2023b,Harikane2022b,Bouwens2023,PerezGonzalez2023,Chemerynska2023,McLeod2024}, the density of galaxies (and in particular of the brightest ones) at $z>9$ is significantly larger than  previously estimated by extrapolation of lower redshift observations as well as theoretical models. Several scenarios have been proposed to explain these findings, ranging from a higher star-formation efficiency, the effect of stochastic star-formation histories, a lower dust extinction, an increased luminosity owing to the contribution of active galactic nuclei (AGNs) or of low-metallicity stars, or even nonstandard cosmological models  \citep[e.g.,][]{Ferrara2022,Ferrara2023,Mason2022,Haslbauer2022,Kohandel2022,Ziparo2022,Fiore2023,Melia2023,Padmanabhan2023,Trinca2024}. 
Follow-up spectroscopy of the newly discovered high-redshift candidates is fundamental both to confirm the measured \textquote{excess} compared to theoretical predictions \citep[e.g.,][]{Harikane2024} and to understand its physical origin. Early spectroscopic campaigns carried out with {\it JWST} NIRSpec have already provided support to the robustness of photometric selections and enabled the exploration of the physical conditions of galaxies at unprecedented redshifts \citep[e.g.,][]{Curtis-Lake2022,ArrabalHaro2023a,ArrabalHaro2023b,Boyett2023,Roberts-Borsani2023,Wang2023}.

Early results have found a trend of decreasing metallicity and increasing excitation and ionization efficiency with increasing redshift  \citep[][]{Trump2023,Tang2023,Nakajima2023,Curti2023}, although most of the sources show physical conditions comparable to those of low-redshift analogs \citep[e.g.,][]{Schaerer2022,Cameron2023a}.
A relatively small number of objects have shown features that are not usually found in low-redshift counterparts and which may be due to physical properties unique to the first phases of star formation and galaxy assembly. A tantalizing example is the bright galaxy GNz11 at $z=10.6$ \citep{Oesch2016}, whose NIRSpec spectrum shows evidence of a nitrogen abundance which is higher than expected for its metallicity \citep{Bunker2023,Cameron2023}. The discovery of other objects with a comparable nitrogen enrichment  \citep{Jones2023,Pascale2023,Isobe2023a,Topping2024} has suggested that we may be witnessing the formation of globular-cluster progenitors \citep{DAntona2023,Bekki2023,Senchyna2023,Marques-Chaves2024,Watanabe2024}. Instead, a high C/O ratio in galaxy GSz12 at $z=12.5$ has been interpreted as the imprint of ejecta from a previous generation of Population~III stars \citep{DEugenio2023}.
A suprisingly large incidence of AGNs has also been suggested by NIRSpec follow-up observations of high-redshift objects \citep[e.g.,][]{Larson2023,Kokorev2023,Fujimoto2023b,Maiolino2023b}, with candidates reaching $z\approx 10$ and beyond \citep[e.g.,][]{Goulding2023,Bogdan2024}, including GNz11 itself \citep{Maiolino2023}.

It is thus fundamental to push spectroscopic investigations to larger samples and higher redshifts to achieve a deeper understanding of the physical conditions of early star-forming regions and to assess the potential contribution of AGN accretion to the ultraviolet (UV) emission of distant galaxies.
 
In this paper, we present the spectroscopic confirmation and characterization of galaxy GHZ2/GLASS-z12\footnote{The two discovery papers appeared on arXiv the same day, and named the galaxy GHZ2 \citep{Castellano2022b} and GLASS-z12 \citep{Naidu2022b}. In the remainder of the paper, we refer to it only as GHZ2 for conciseness.}, initially discovered as a robust $z \approx 12.0$--12.5 candidate by \citet[][C22 hereafter]{Castellano2022b} and \citet{Naidu2022b} in the GLASS-JWST Early Release Science NIRCam field \citep{Treu2022} \citep[see also][]{Bouwens2022,Donnan2023,Harikane2022b,Atek2023}. GHZ2 provided the first example of an unexpected population of high-redshift bright galaxies and was targeted with {\it JWST} NIRSpec multi-object spectroscopy through Program GO-3073 (PI M. Castellano), which is aimed at extensive follow-up observations of the $z\gtrsim 10$ candidates selected by C22 and \citet[][]{Castellano2023a} in the GLASS-JWST region. A companion paper presents JWST MIRI  spectroscopy of GHZ2 at $\lambda>$5$\mu$m under program GO-3703 (PI J. Zavala) \citep{Zavala2024}. A combined analysis of the two datasets is presented in \citet{Calabro2024}.

The paper is organized as follows. We describe observations and data reduction in Sec.~\ref{sec:obs} and discuss the main features detected in the NIRSpec spectrum in Sec.~\ref{sec:spectrum}. In Sec.~\ref{sec:SFRAGN} we investigate the source of ionizing photons by comparing GHZ2 to models of AGN and star-formation emission.  Sec.~\ref{sec:SFSCENARIO} presents the properties of the object and a potential physical scenario for GHZ2 in the context of high-redshift star formation. We summarize the results and discuss future prospects in Sec.~\ref{sec:summary}. 

Throughout the paper we adopt AB magnitudes \citep{Oke1983}, a \citet{Chabrier2003} initial mass function (IMF), a solar metallicity of 12 + log(O/H) = 8.69 \citep{Asplund2009},  and a flat $\Lambda$CDM concordance model (H$_0$ = 70.0~km~s$^{-1}$~Mpc$^{-1}$, $\Omega_M=0.30$). 

\begin{figure*}[ht]
\centering
\includegraphics[trim={1.8cm 2.0cm 1.8cm 2.5cm},clip,width=18cm,keepaspectratio]{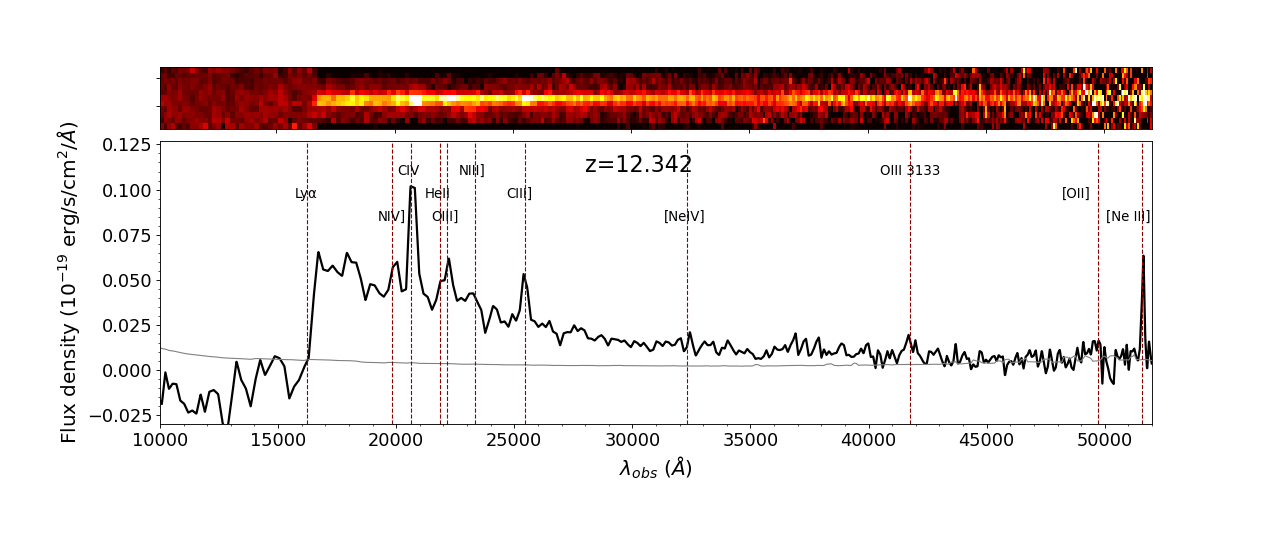}
\caption{Observed 2D (top) and 1D (bottom) NIRSpec PRISM spectra of GHZ2. In the bottom panel the gray line shows the noise RMS, and red dashed lines highlight the wavelength of the UV features discussed in the present paper.} 
\label{fig_FULLSPEC}
\end{figure*} 

\section{Observations and Data Reduction}\label{sec:obs}

Program GO-3073 comprises two partially overlapping NIRSpec PRISM pointings on the GLASS-JWST NIRCam field. The observation of each pointing is divided into three visits to enable six-band coordinated parallel NIRCam observations on flanking fields. Each visit has an exposure time of 6567~s and adopts a NRSIRS2 readout pattern, standard 3-shutter ``slits'' for the primary targets, and a 3-point nodding. The first pointing was observed on 2023 Oct. 24, with the second pointing being scheduled for 2024 June/July. Unfortunately, two of the three nodding positions of the third visit were affected by an electric short and are unusable. The present paper thus exploits 7 dithered observations of GHZ2 for a total observing time of 15323~s. A forthcoming paper (Napolitano et al., in prep.) will present spectroscopy of the other $z \gtrsim 10$ candidates already observed. 

The data were reduced as outlined by \cite{ArrabalHaro2023a} and \cite{ArrabalHaro2023b} with the STScI Calibration Pipeline\footnote{\url{https://jwst-pipeline.readthedocs.io/en/latest/index.html}} version 1.13.4. We provide here a brief description of the main steps. The pipeline modules are divided into three components. In summary, the \textsc{calwebb\_detector1} module corrects for detector $1/f$ noise, subtracts dark current and bias, and generates count-rate maps (CRMs) from the uncalibrated images. The \textsc{calwebb\_spec2} module creates two-dimensional (2D) cutouts of the slitlets, corrects for flat-fielding, performs background subtraction using the three-nod pattern, executes photometric and wavelength calibrations, and resamples the 2D spectra to correct distortions of the spectral trace. The \textsc{calwebb\_spec3} module combines images from the three nods, utilizing customized extraction apertures to extract the one-dimensional (1D) spectra.

We found that 2 out of 7 dithered observations of GHZ2 from 2 different visits include a secondary, low-redshift (z=1.68) object in the upper part of the slit. When performing the standard background subtraction, the H$\alpha$ line of the contaminating source would create a spurious absorption feature in the main target spectrum at $\sim 17,500$--17,700~\AA. We therefore masked the contaminated 2D regions before applying the three-nod pattern background subtraction. Also, since the third visit has only one usable dither, we applied a custom master background subtraction by defining local  {\it ad hoc} background windows in the adjacent empty shutters. 
Subsequently, both 2D and 1D spectra were examined with the \textsc{Mosviz} visualization tool \citep{Developers2023} to mask potential remaining hot pixels and artifacts in the spectra. Following the masking of image artifacts, data from three consecutive exposure sequences were consolidated to produce the final 2D and 1D spectral products. No contamination due to the secondary source is apparent in the final science spectrum. Nonetheless, we will not consider in the following analysis the region of the H$\alpha$ line of the secondary object which was found to severely affect the GHZ2 spectrum at 1310--1330~\AA\ rest-frame when performing a standard nodded background correction. 

We correct for wavelength-dependent slit and aperture losses by matching the spectrum to the NIRCam broadband photometry as follows. Synthethic photometry is first computed by integrating the spectrum with the relevant transmission curves in the F200W, F277W, F356W, and F444W bands.  We then fit linear relations to both the original NIRCam  and the synthetic NIRSpec photometry, and use the wavelength-dependent ratio of the two to correct the spectrum longward of the Lyman break. The correction ranges from 1.19 at 2~$\mu$m to 1.78 at 5~$\mu$m. These values are consistent with the correction factors found for prism spectra of bright z$>$10 galaxies by \citet{ArrabalHaro2023b}.

In this work we use an updated measurement of the NIRCam photometry in the A2744 field which will be discussed in detail in a forthcoming paper (Merlin et al., in prep.). The new measurements exploit the latest reduction of the GLASS-JWST NIRCam data including new observations acquired in July 2023. As discussed by C22, and stated above, GHZ2 is close ($\sim 0.5''$) to a foreground galaxy that also contaminates the total flux measured within the Kron ellipse \citep{Kron1980}. We thus estimated its total flux with the \textsc{T-PHOT} software \citep[][]{Merlin2015,Merlin2016a} using the GHZ2 light profile in the F277W band as high resolution prior to re-extracting the F444W photometry, finding a total magnitude of $27.05 \pm 0.02$. Fluxes in the other bands were measured by scaling the aforementioned total flux according to colors measured on point-spread-function (PSF) matched images with \textsc{A-PHOT} \citep{Merlin2019} in an aperture with a diameter two times the PSF full width at half-maximum intensity (FWHM = $0\farcs28$) \citep[see also][]{Merlin2022,Paris2023}. 

Finally, throughout the work we will derive rest-frame physical properties taking into account that GHZ2 is affected by moderate lensing magnification ($\mu = 1.3$) estimated on the basis of the model by \citet{Bergamini2023}. Line ratios and equivalent widths (EWs) are unaffected by lensing.

\section{UV Spectrum of GHZ2}\label{sec:spectrum}

\begin{figure*}
\centering
\includegraphics[trim={0.cm 0.8cm 0.2cm 1.5cm},clip,width=5cm,keepaspectratio]{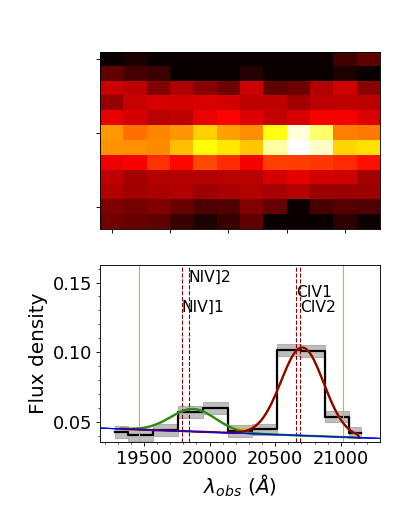}
\includegraphics[trim={0.cm 0.8cm 0.2cm 1.5cm},clip,width=5cm,keepaspectratio]{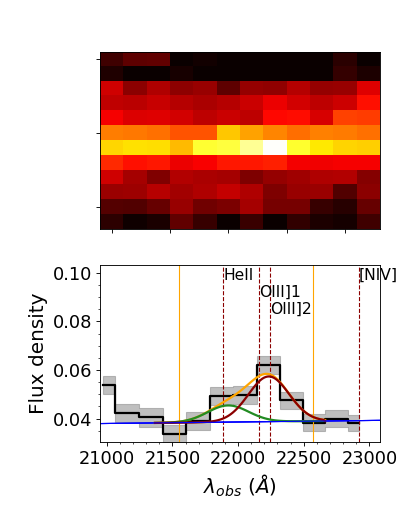}
\includegraphics[trim={0.cm 0.8cm 0.2cm 1.5cm},clip,width=5cm,keepaspectratio]{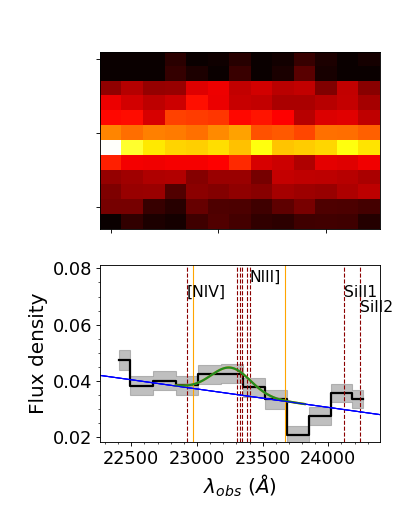}
\includegraphics[trim={0.cm 0.8cm 0.2cm 1.5cm},clip,width=5cm,keepaspectratio]{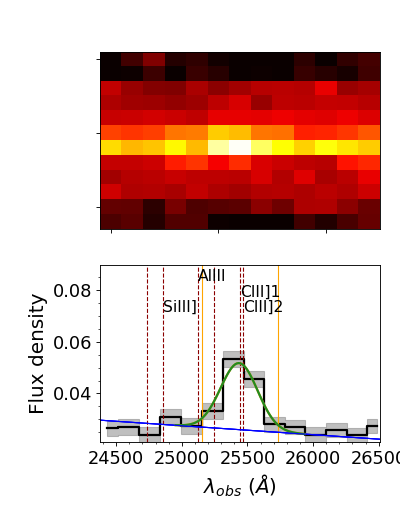}
\includegraphics[trim={0.cm 0.8cm 0.2cm 1.5cm},clip,width=5cm,keepaspectratio]{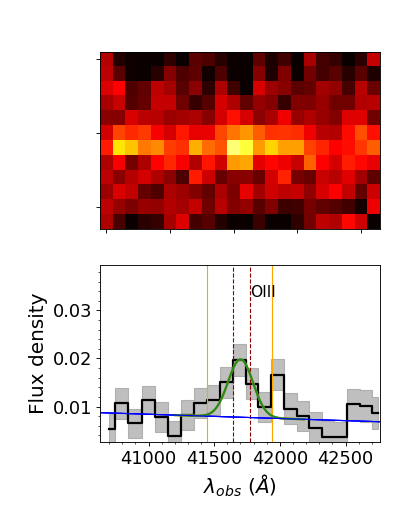}
\includegraphics[trim={0.cm 0.8cm 0.2cm 1.5cm},clip,width=5cm,keepaspectratio]{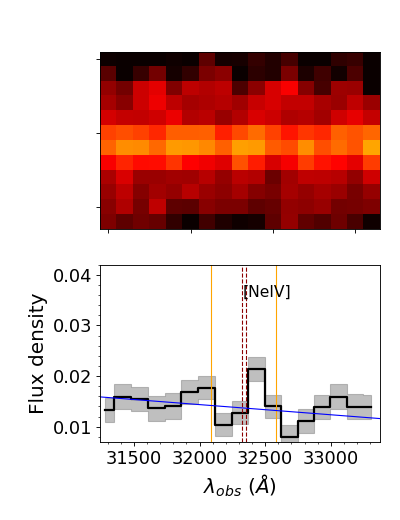}
\includegraphics[trim={0.cm 0.8cm 0.2cm 1.5cm},clip,width=5cm,keepaspectratio]{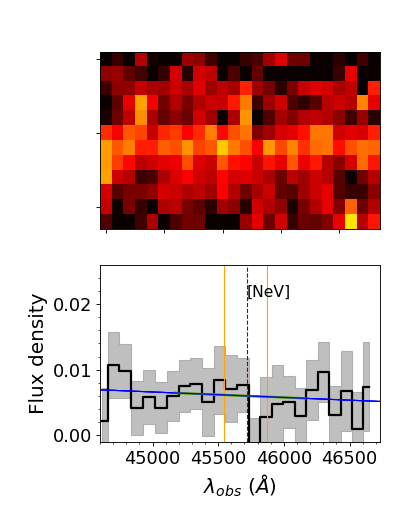}
\includegraphics[trim={0.cm 0.8cm 0.2cm 1.5cm},clip,width=5cm,keepaspectratio]{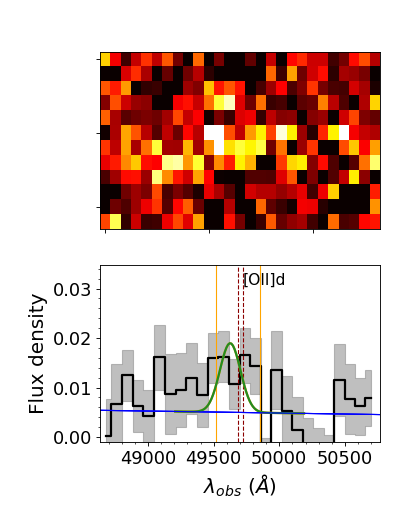}
\includegraphics[trim={0.cm 0.8cm 0.2cm 1.5cm},clip,width=5cm,keepaspectratio]{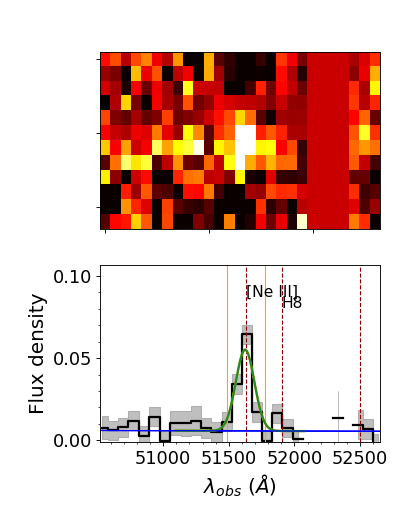}
\caption{From top left to bottom right: snapshots of the NIRSpec spectrum in regions with width of 160~\AA\ rest-frame centered at the position of \civ\ and \niv; \heii\ and \oiii; the tentative \niii\ line; \ciii; \oiiibowen, \neiv; \nev; \oii\ and \neiii. All flux densities are in units of 10$^{-19}$ \ergsA. The gray shaded area shows the 1$\sigma$ uncertainty in each pixel. Red dashed lines indicate the wavelength of all potential features in the relevant spectral range. The vertical orange lines enclose the region where the signal-to-noise ratio (SNR) of the feature is evaluated from direct integration. For all significant lines the relevant single-Gaussian fit is shown in green. When a double-Gaussian fit was used, the two components are shown as green and red curves, and the sum of the two in orange. The blue line in each panel shows the estimated UV continuum. } 
\label{fig_LINES}
\end{figure*} 

\subsection{Redshift determination, emission-line detection and flux measurement}\label{subsec:redshiftofGHZ2}
The NIRSpec spectrum of GHZ2 (Fig.~\ref{fig_FULLSPEC}) shows a sharp Lyman break and clear emission features consistent with $z \approx 12.3$. We obtain a first measurement of its redshift ($z= 12.33 \pm 0.02$) from the centroid of the \neiii\ emission which is the single line measured at the highest spectral resolution ($R=300$) among those detected at high SNR. 

To measure the flux and EW of all the emission lines, we  perform a direct integration of the continuum-subtracted spectrum in a window centred at the expected wavelength $\lambda$ and having a width $4\times \sigma_R (\lambda)$, where $\sigma_R (\lambda)$ is the expected Gaussian root-mean-square (RMS) of a line observed at resolution $R(\lambda$). In the case of partially blended lines we assess the significance of the entire line complex and of the different components in narrower windows with width $2 \times \sigma_R (\lambda)$. The continuum is measured as a linear interpolation of the regions free of potential features closest to each line. To make sure that the examined lines are not affected by unmasked artefacts, we inspected all the single-dithered spectra and the spectra obtained by separately combining the two visits that were observed at three nodding positions. We consider as significant every detection with SNR $>5$, where the uncertainty takes into account errors in both the integrated flux and the extrapolated continuum at the line position. 

A Gaussian fit of the continuum-subtracted flux is then performed for all significant emission features using the \texttt{specutils} package of \texttt{astropy}. Unresolved doublets and multiplets are treated as a single Gaussian profile, while a double-Gaussian fit is used for partially blended lines (see Fig.~\ref{fig_LINES}). We let the mean of the Gaussian vary according to a $\Delta z=0.04$ and the Gaussian RMS within 5\% of the nominal $\sigma_R (\lambda)$ to account for the uncertainty in the redshift and considering the potential uncertainty in the centroid and RMS of unresolved multiplets. 
We report in Table~\ref{linestable} the measured fluxes, EWs, and the significance of the lines obtained from the direct integration test described above. The uncertainties are obtained through a Monte Carlo simulation taking into account errors in both the Gaussian fit and the continuum, and are thus more conservative than the SNR measured through direct integration. In the case of not significant lines, the 3$\sigma$ limits obtained through direct integration are reported. 

All lines yield redshift estimates that are consistent within the relevant uncertainty with the value obtained from the \neiii\ line. From a weighted average of the measurements of the best resolved, high-SNR lines (\niv, \civ, \ciii, and \neiii), we obtain $z=12.342 \pm 0.009$, which we adopt  hereafter.  

\subsection{UV emission lines}
\label{subsec:detectedlines}
The most prominent line is the unresolved \civd\ doublet\footnote{In Table~\ref{linestable} and in the discussion we will refer to unresolved doublets and multiplets as a single line at the relevant central wavelength.} which is at close separation from the unresolved \nivd\ doublet. We measured the C~IV + N~IV complex with a double-Gaussian fit as described above, obtaining  EW(C~IV) = 46~\AA~and EW(N~IV)= 12~\AA.
This EW is $\sim$20 times higher than the average value found in Ly-$alpha$ emitting systems at z=5-11 \citep{RobertsBorsani2024} and places GHZ2 in the class of high-redshift C~IV emitters such as RXCJ2248-ID at $z=6.11$ \citep[EW(CIV)$\approx$34\AA, ][]{Balestra2013,Schmidt2017, Mainali2017,Topping2024} and A1703-zd6 at $z=7.045$ \citep[EW(CIV)$\approx$38\AA, ][]{Stark2015b}. We have checked the spatial extent of the C~IV line to search for evidence of outflows associated to this strong emission feature. We compared the spatial profile of the line, averaged over a wavelength window with width $2\times \sigma_R (\lambda)$, to the spatial profile of the continuum averaged in the two closest windows free of emission features. We find that the line and continuum profiles are statistically consistent, providing no indication of spatially extended C~IV outflows. In Appendix~\ref{appendix:XSHOT}, we discuss the detection of the C~IV $\lambda 1548$ line in a VLT X-SHOOTER spectrum, suggesting that its flux is mostly attributed to a component with FWHM$\approx$ 390 $\pm$ 160~km~s$^{-1}$. Similarly, we detected the blended O~III + He~II complex and separated the two components with a double-Gaussian fit, finding an EW(\heii)=4.9$\pm$3.1~\AA, comparable to the highest values found at low/intermediate redshifts \citep[e.g.,][]{Berg2019,Nanayakkara2019,Saxena2020}. The \oiiit\ doublet has a total EW $\approx 14$~\AA, and it is $\sim 2.7$ times brighter than \oiid\ which is only marginally detected. The ratio between the two oxygen lines, as well as the presence of a prominent \neiii\ line, points to a high ionization and low metallicity, as we will discuss in more detail in Sec.~\ref{sec:met}. The \ciii\ line is also found at high significance with  EW $\approx 25$ \AA, at the extreme end of the range measured at high redshift \citep[e.g.,][]{LeFevre2019,Llerena2021}. GHZ2 lies significantly above the redshift-EW(CIII) relation derived by \citet{RobertsBorsani2024} for the general high-z population (13.7$\pm$0.9 \AA~at z$\sim$11 and 17.3 \AA~inferred for z=12.34), similarly to the z=12.5 galaxy GSz12 \citep{DEugenio2023}. 

We did not significantly detect the very high-ionization lines \neiv\ and \nev, with 3$\sigma$ upper limits on their flux of $\sim 2.2 \times 10^{-19}$ and $\sim 3.1 \times 10^{-19}$~erg~s$^{-1}$~cm$^{-2}$, respectively.

\subsection{Tentative detection of the N~III] $\lambda1750$ line}\label{subsec:NIII}
We obtained a 7$\sigma$ detection of the (unresolved) \niii\ multiplet. This line is extremely important to assess the N/O ratio (Sec.~\ref{sec:met}) and whether GHZ2 belongs to the class of N-enhanced objects discussed in the recent literature \citep[e.g.,][]{Bunker2023,Jones2023,Topping2024}. 
However, we note that the SNR of this detection is heavily dependent on the estimate of the local continuum, which is challenging to measure at the resolution of the prism in the region of the N~III] line. In the baseline measurement of the line we used the closest windows that are distant enough from potential features, at 1690--1710~\AA\ and 1840--1850~\AA\ rest-frame. As an alternative, we estimated a global continuum as a third-degree polynomial fit of all regions in the GHZ2 spectrum that are at a distance in wavelength of at least 2$\sigma(R)$ from any potential emission or absorption line. We found that our conclusions remain unchanged for all the lines discussed in the previous section and, most importantly, for the \niii\ line. However, the detection is found not to be significant, with a 3$\sigma$ upper limit of 0.9 \ergs, if we adopt as an estimate of the continuum the median value in the blueward window at 1690--1710~\AA. While this is not unexpected considering the steep shape of the UV continuum of GHZ2, it reflects the tentative nature of the detection; future deeper/higher resolution observations are needed to confirm it. In Sec.~\ref{sec:met}, we will address the implications of both the case of a detection and of the conservative upper limit mentioned above.

\begin{deluxetable}{cccc}\label{linestable}
\tablecaption{UV emission lines in GHZ2}
\tablewidth{0pt}
\tablehead{
\colhead{Line} & \colhead{Flux} & \colhead{EW} & \colhead{SNR} \\
\colhead{} & \colhead{($10^{-19}$~erg~s$^{-1}$~cm$^{-2}$)} & \colhead{(\AA)} &
}
\startdata
Ly$\alpha$ & $<$8.5  & $<$10 & $<$3\\
\niv & 6.9 $\pm$  0.6 & 12.1  $\pm$    1.2 & 13  \\
\civ & 25.7   $\pm$   0.6  & 45.8   $\pm$   1.2 & 43 \\
\heii & 2.7   $\pm$   1.6  & 4.9  $\pm$    3.1 & 5 \\
\oiii & 7.2  $\pm$    1.5  & 13.7  $\pm$    4.8 & 9\\
\niii & 3.4  $\pm$    0.9  & 6.8  $\pm$    2.1 & 7 \\
\ciii & 9.1  $\pm$    0.2  & 25.6  $\pm$    12.5 & 11\\
\neiv & $<$2.2 & $<$16 & $<$3\\
\oiiibowen & 2.9  $\pm$    0.5  & 30.5  $\pm$    10.3 & 9\\
\nev &  $<$3.1  & $<$75 & $<$3\\
\oii & 2.7  $\pm$   1.1  & 42  $\pm$   28 & 12 \\
\neiii & 6.4   $\pm$   0.8  & 35.5   $\pm$   4.2 & 32\\
\enddata
\tablecomments{The SNR is evaluated from the integration of the continuum-subtracted flux in a region centered on each feature; fluxes and EWs are measured with Gaussian fits. See Sect.~\ref{subsec:redshiftofGHZ2} for details. All upper limits are at 3$\sigma$.}
\end{deluxetable}

\subsection{Detection of the O~III $\lambda$3133 Bowen fluorescence line}\label{subsec:OIIIbowen}

We significantly detected (SNR $\approx 9$) an emission line at $\lambda_{\rm obs}=4.17~\mu$m, corresponding to $\sim 3128$~\AA\ rest-frame. The only known features compatible with the measured wavelength are the O~III lines at 3123~\AA\ and 3133~\AA\ emitted via Bowen  resonance fluorescence \citep{Bowen1934,Bowen1947}, no other potential emission lines can affect the examined wavelength range. These O~III emission lines arise in a highly ionized and dense environment (though in the case of GHZ2, not so dense as to inhibit the formation of forbidden lines), with strong UV ionizing flux. Briefly, He~II Ly$\alpha$ photons emitted by the ionized gas at rest-wavelength 303.782~\AA\ can be absorbed by O~III at a nearly coincident wavelength (303.800~\AA), exciting its 2p3d$^3$P$_2$ level. When the O~III optical depth is large, the  He~II Ly$\alpha$ photons find a way out of resonance through a cascade of fluorescent transitions in the optical and UV, including the 3123 and 3133~\AA\ lines.
The O~III line at 3133~\AA\ is the brightest and most commonly observed Bowen transition of this kind, although the resolution of the GHZ2 spectrum does not allow us to exclude some contribution from the (generally much weaker) O~III Bowen transition at 3123~\AA\ in the detected emission.  For a detailed discussion of the physics of these transitions in different astrophysical contexts, see  \citet{Harrington1972}, \citet{Deguchi1985}, and \citet{Djenivze2003}.  

The Bowen lines are usually observed in symbiotic and X-ray binaries \citep[e.g.,][]{Schachter1989,Schachter1991,Pereira1999,Selvelli2007}, planetary nebulae \citep[e.g.,][]{Liu1993}, and some low-redshift Seyfert galaxies \citep[e.g.,][]{Malkan1986,Schachter1990}. Nearly all Seyfert 2 nuclei exhibit Bowen lines, at least weak ones, if the samples of \cite{Malkan1986} and \cite {Schachter1990} are representative; however, 
although Seyfert 1 nuclei and quasars should also have Bowen lines \citep{Netzer1985}, those produced by the broad-line region are much more difficult to detect and quantify because they have low EW and merge with the continuum or with other lines (Fe~II, etc.).  While fluorescent transitions have been previously observed at intermediate redshifts thanks to strong gravitational lensing \citep{Vanzella2020}, to our knowledge GHZ2 is the first reionization-era object showing the \oiiibowen~emission feature.

The \oiiibowen\ line might suggest that GHZ2 is an AGN. However, the nondetection of the [Ne~V] line is puzzling; this line is generally 2--5 times brighter than the O~III line in AGNs \citep{Malkan1986,Schachter1990,Kraemer1994,Lanzuisi2015}, with the only known exceptions being Mrk~42 \citep{Malkan1986} and events belonging to the class of \textquote{Bowen fluorescence flares} interpreted as enhanced accretion episodes onto an already active super-massive black hole \citep[][]{Trakhtenbrot2019,Makrygianni2023}. 
Since the ionization potentials of Ne$^{2+}$, Ne$^{3+}$, and Ne$^{4+}$ are (respectively) 63.45, 97.12, and 126.2~eV, and only [Ne~III] $\lambda$3868 is visible, it is likely that the highest-energy ionizing photons are 70--90~eV. 
A low abundance of Ne is also possible \citep[e.g.,][]{Isobe2023b}, given that [Ne~III] $\lambda$3868 is generally $\sim 10$ times stronger than \oiiibowen\ in AGNs yet is only about twice as strong in GHZ2.

It is perhaps also curious that there is no sign of He~II $\lambda 3204$ in the spectrum of GHZ2. This line is typically $\sim 60$\% the strength of \oiiibowen\ in Seyfert nuclei \citep[][]{Malkan1986,Schachter1990}, so it should be detectable, albeit at low SNR. However, its strength can sometimes be as low as $\sim 25$\% that of \oiiibowen;
if this is the case in GHZ2, then it would not be detected in the current data. Moreover, based on Tables 2 and 3 of \citet{Seaton1978}, He~II $\lambda 3204$ is usually $\sim 0.06$ times the strength of He~II $\lambda 1640$ in gases with $T =$ (1--2) $\times 10^4$~K and $n_e = 10^4$--$10^6$~cm$^{-3}$; given the measured flux of He~II $\lambda 1640$ in the spectrum of GHZ2, it thus might not be surprising that He~II $\lambda 3204$ is undetected.

A higher-SNR spectrum of GHZ2 is needed to detect and measure accurately He~II $\lambda 3204$, and thus be able to determine the Bowen yield ($y_{\rm HeO}$, the fraction of He~II Ly$\alpha$ converted to O~III Bowen lines) from the relative intensity of the two lines; the relevant relation \citep{Schachter1990} is $y_{\rm HeO} = 0.12(I(3133)/I(3204))$.
We can also use He~II $\lambda 1640$ to estimate the Bowen yield;
$y_{\rm HeO} = 0.12[I(3133)/I(3204)] = 0.12[I(3133)/I(1640)][I(1640)/I(3204)]$ and (as noted above) $I(1640)/I(3204) \approx 0.06$. 
However, the measured intensity ratio $I(3133)/I(1640)$ in GHZ2 is $1.1 \pm 0.8$, quite uncertain. Our formal result is that $y_{\rm HeO} = 3.8 \pm 2.9$, where the error bar includes only the uncertainty in $I(3133)/I(1640)$, not the variation in $I(1640)/I(3204)$ among gases having different physical conditions. A value  $\ge 1$ is unphysical, but in any case our result is so uncertain that the Bowen yield is essentially unconstrained. 
Moreover, He~II $\lambda$3204 line is preferred over He~II $\lambda$1640 when determining the Bowen yield, even though the latter is intrinsically a factor of 16--17 stronger than the former, since the result with He~II $\lambda$3204 is independent of reddening. 

\begin{figure*}
\centering
\includegraphics[trim={0.8cm 0.2cm 0.8cm 0.5cm},clip,height=8cm,keepaspectratio]{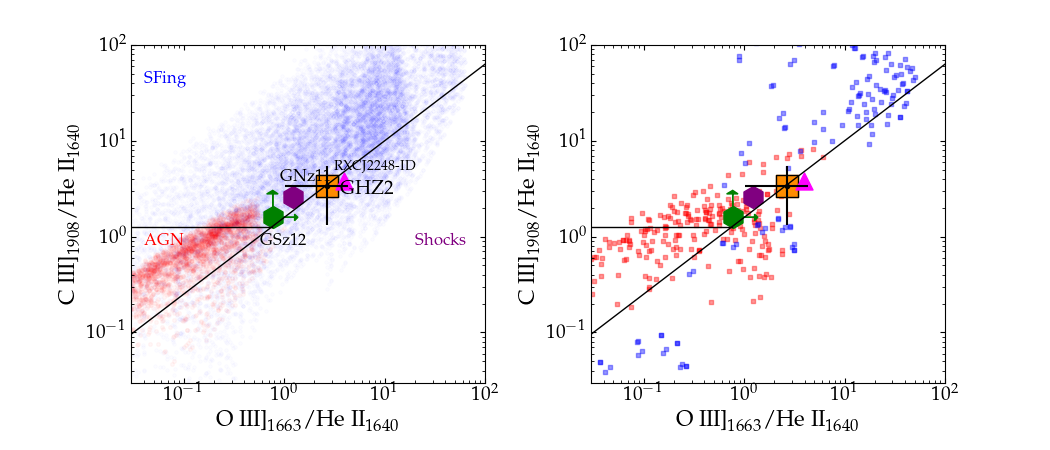}
\includegraphics[trim={0.8cm 0.2cm 0.8cm 0.5cm},clip,height=8cm,keepaspectratio]{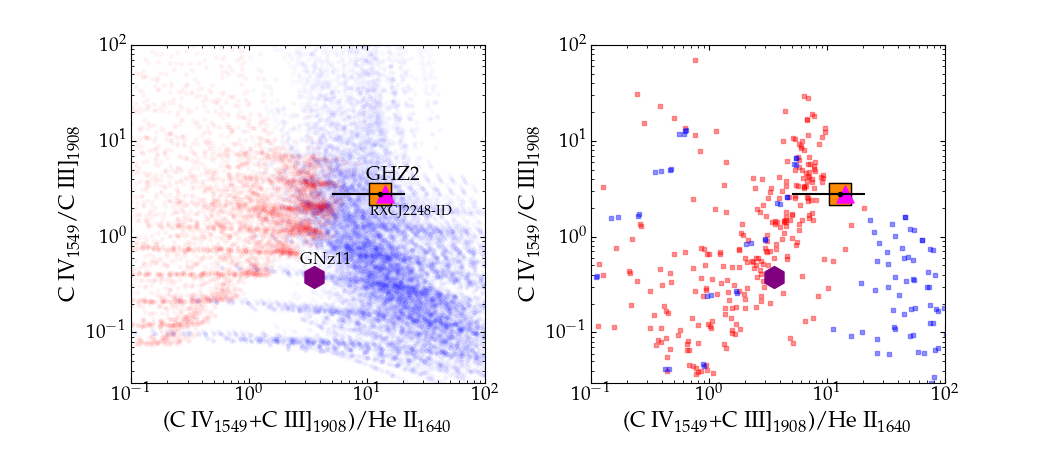}
\caption{The position of GHZ2 (orange-filled square with black error bars) in the UV line diagnostic diagrams discriminating between star formation and an AGN as the main ionizing source: C~III]/He~II versus O~III]/He~II \citep[top, the black lines divide the diagram according to the selection criteria from][]{Mingozzi2023} and C~IV/C~III] versus (C~IV+C~III])/He~II (bottom). The AGN and star-forming models from \citet{Feltre2016} and \citet{Gutkin2016} (left panels) and from \citet{Nakajima2022} (right panels) are shown in red and blue, respectively. The positions of reference objects are shown when relevant data are available: GNz11 (purple hexagon), GSz12 (green hexagon), and \rxc\ (magenta triangle) using measurements from \citet{Bunker2023}, \citet{DEugenio2023}, and \citet{Topping2024}, respectively.} 
\label{fig_SFAGN1}
\end{figure*} 

\begin{figure*}
\centering
\includegraphics[trim={1.8cm 2.5cm 1.8cm 0.5cm},clip,height=16cm,keepaspectratio]{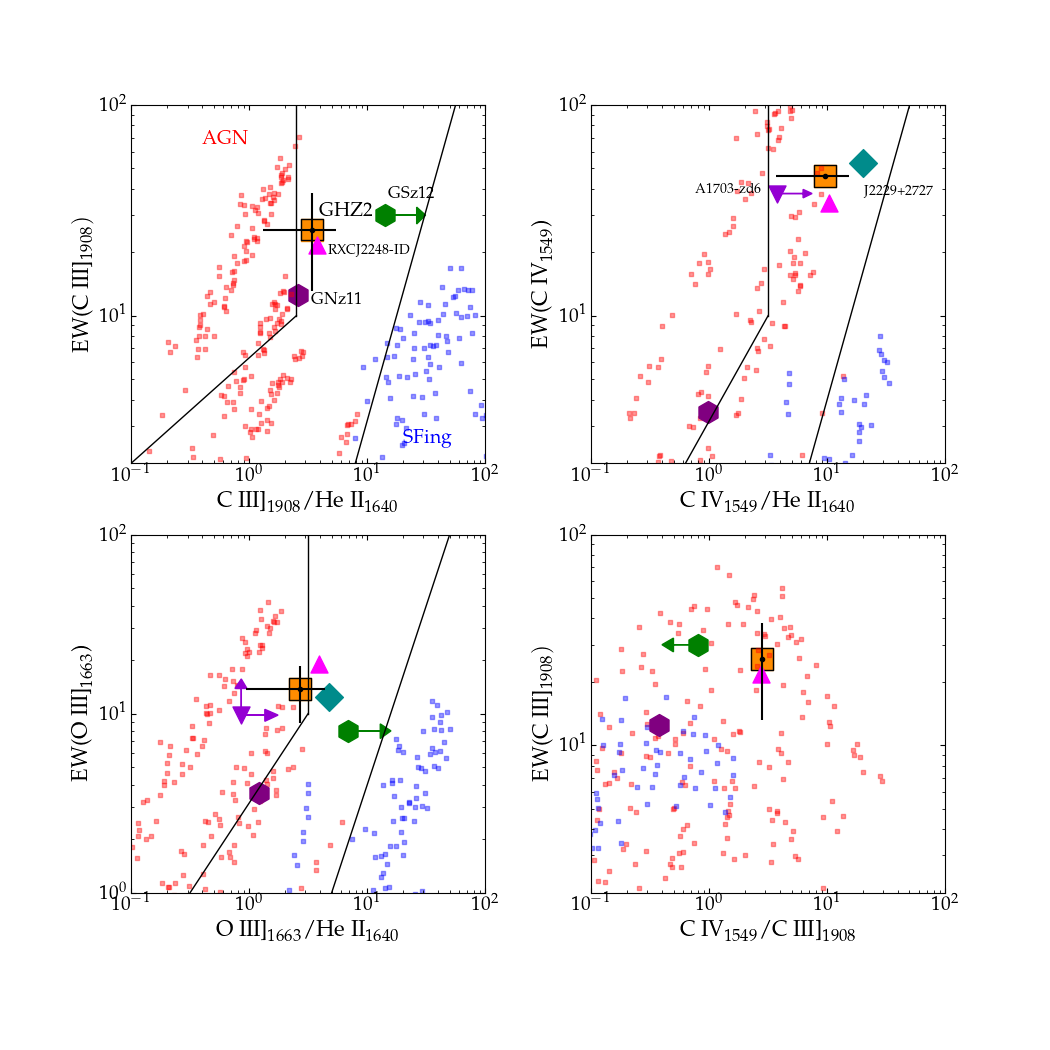}
\caption{Diagnostic diagrams based on the EW of UV lines including, where available, the selection regions from \citet{Hirschmann2019}. Models are from \citet{Nakajima2022}. The comparison with observed sources is as in Fig.~\ref{fig_SFAGN1}, but including in the top-right and bottom-left panels here also objects  \JIzotov\ \citep[dark cyan diamond;][]{Izotov2024} and  \AStark\ \citep[purple triangle;][]{Stark2015b}. The EW(O~III]) for object \AStark\ is considered an upper limit because only one of the two components of the doublet is measured.} 
\label{fig_SFAGN2}
\end{figure*} 

\begin{figure*}
\centering
\includegraphics[trim={1.5cm 0.2cm 2.5cm 0.5cm},clip,width=16cm,keepaspectratio]{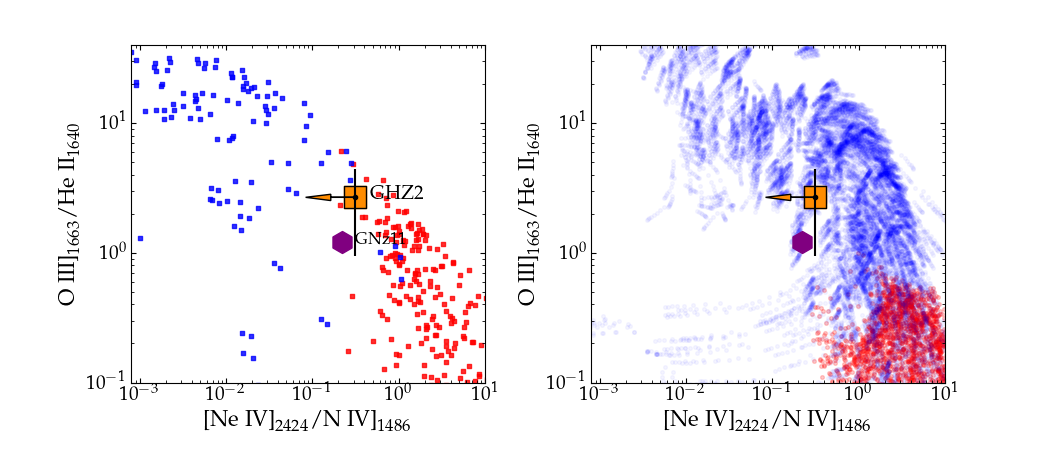}
\caption{The position of GHZ2 and GNz11 in the O~III]/He~II versus [Ne~IV]/N~IV] plane compared with models from \citet{Feltre2016} and \citet{Gutkin2016} (left panel) and from \citet{Nakajima2022} (right panel). Symbols as in Fig.~\ref{fig_SFAGN1}.} 
\label{fig_SFAGN_highion}
\end{figure*} 

\subsection{UV continuum and limits on Ly$\alpha$ emission}\label{subsec:lya}
Finally, we exploited the continuum measured in the NIRSpec spectrum to constrain the UV magnitude and slope of GHZ2 free of the systematics introduced by emission lines. The UV slope $\beta$ was measured by fitting the continuum flux with a power law after masking all potential emission features. While the fit is usually performed at 1400--2600~\AA\ rest-frame \citep{Calzetti2000}, we restrict our analysis to the 1400--2600~\AA\ range because the potential impact of damped Ly$\alpha$ absorption (DLA) wings can affect the continuum at shorter wavelengths \citep{Heintz2024}.  We employed \textsc{emcee} \citep{Foreman_Mackey2013} to perform a Markov Chain Monte Carlo (MCMC) analysis identifying the best-fitting model free parameters through 100 chains and 100,000 steps. As a prior, we imposed a flat distribution $-3 \leq \beta \leq 0$. A UV slope of $\beta$= -2.46 $\pm$ 0.08 was measured, where the best-fit value and its uncertainty are the median and standard deviation of the posterior distribution. In this respect, GHZ2 is consistent with the most UV-luminous and high stellar mass sources at z$>$6, which show little evolution between z$\sim$6-11 ($\beta\sim$-2.4 to -2.5), indicative of mostly dust-poor systems \citep{RobertsBorsani2024}.
We then derived a UV absolute magnitude $M_{\mathrm{UV}}$ = -20.53 $\pm$ 0.01  from the average continuum flux of the fitted power law in the range 1450--1550~\AA\ rest-frame, after correcting for magnification. 
Interestingly, we obtain a bluer UV slope $\beta= -2.54 \pm 0.07$ by fitting the NIRCam photometry in the F200W, F277W and F356W bands due to the impact of the emission lines at 2$\mu$m.

The measurement of Ly$\alpha$ emission in PRISM spectra is challenging owing to the poor resolution and to the effects of interstellar and intergalactic absorption at the Lyman break.

We followed the procedure described in detail by \cite{Jones2023C} and \cite{Napolitano2024} to derive a limit on the rest-frame Ly$\alpha$ EW on the basis of the measured redshift, the resolution of the instrument at the Ly$\alpha$ wavelength, and the continuum flux and uncertainty redward of Ly$\alpha$. The continuum is extrapolated on the basis of the UV slope measured as described above. We derived a 3$\sigma$ limit of EW(Ly$\alpha$) $ < 10$~\AA. A consistent limit is found when using the continuum estimated using a third-degree polynomial fit of line-free regions (Sec.~\ref{subsec:NIII}). 

\section{Source of Ionizing Photons: Star Formation or AGN?}\label{sec:SFRAGN}
The prominent emission lines detected in the GHZ2 spectrum, including high-ionization transitions such as C~IV and N~IV], clearly suggest the presence of a hard ionizing source. In order to discriminate whether the dominant ionization source of GHZ2 is an AGN or emission from young stellar populations, we tested several diagnostics exploiting ratios and EWs of UV lines \citep[e.g.,][]{Feltre2016,Nakajima2018a,Hirschmann2019,Hirschmann2023,Mingozzi2023}. In the following, we compare GHZ2 to two different sets of models built by processing stellar or AGN emission with the photoionization code \textsc{CLOUDY} \citep[][]{Ferland2013}. We consider models for star-forming galaxies by \citet[][G16 hereafter]{Gutkin2016} based on the most recent version of the \citet[][]{Bruzual2003} stellar population synthesis models \citep[see][for details]{VidalGarcia2017}, and the narrow-line AGN models by \citet[][F16 hereafter]{Feltre2016}. We also compare with Pop~II star-forming models \citep[based on BPASS v2.2.1 SEDs,][]{Stanway2018} and AGN models by \citet[][NM22 hereafter]{Nakajima2022}. We refer the reader to the above-mentioned papers for a detailed description of the modeling procedures and assumptions. While the considered models encompass a large range of values for metallicity, ionization parameter, and (in the case of F16 and G16) also C/O ratio and gas density, we chose not to restrict the comparison to any predefined range of values considering the unknown nature of GHZ2.  

When possible, we compare GHZ2 to reference objects with well-characterized spectra: GNz11 \citep[$z=10.6$;][]{Bunker2023,Maiolino2023}, GSz12 \citep[$z=12.48$;][]{DEugenio2023}, the high-redshift C~IV emitters with EW$>$30\AA~RXCJ2248-ID \citep[$z=6.1$;][]{Topping2024} and A1703-zd6 \citep[$z=7.045$;][]{Stark2015b}, and the only similar example known at low redshift, galaxy J2229+2727 \citep[$z=0.07622$;][]{Izotov2024}. 

For the comparison  we exploit a wide range of emission-line ratios that are reported in Fig.~\ref{fig_SFAGN1}. In all these cases, GHZ2 usually falls in the region where star-forming and AGN models tend to overlap. Notably, the assessment is extremely model-dependent. GHZ2 is more compatible with star formation according to the models by F16 and G16 in both the C~III]/He~II versus O~III]/He~II and the C~IV/C~III] versus (C~IV+C~III])/He~II diagrams. Instead, the AGN models by NM22 extend to the region where GHZ2 is found. The position of GHZ2 in the C~III]/He~II versus O~III]/He~II diagram is also compatible with emission by shocks according to the selection criteria by \citet{Mingozzi2023}. A comparison with other high-redshift objects with prominent emission lines shows a remarkable similarity between GHZ2 and RXCJ2248-ID in all diagrams. GHZ2 is also very close to GNz11 in the C~III]/He~II versus O~III]/He~II plot, while the much weaker \civ\ emission in GNz11 differentiates the two sources in the other diagrams.Perhaps the ionizing continuum is weaker in GNz11, which could also help explain the absence of obvious Bowen fluorescence lines in its spectrum. We find that both in the case of F16/G16 and NM22 models, the line ratios measured in GHZ2 are typical of low metallicities ($Z/{\rm Z}_{\odot} \lesssim 0.2$) and a high ionization parameter both in the star-formation and AGN case. In fact, differences between the two model sets can be likely explained by the NM22 models allowing for a ionization parameter as high as log\,$U = -0.5$, while the F16 and G16 ones are limited to log\,$U \leq - 1$. On the basis of these comparisons, and considering the significant differences between the various models, we conclude that it is hard to conclusively identify the source of ionizing radiation based only on the line ratios.

We show in Fig.~\ref{fig_SFAGN2} the diagnostic diagrams based on the EWs of the UV lines proposed by \citet{Nakajima2018a} compared to the NM22 models (the only ones that provide EW information). The rest-frame EWs of GHZ2 UV lines are always consistent with AGN emission or with composite emission from both star formation and accretion. GHZ2 would be classified as an AGN according to the thresholds proposed by \citet{Nakajima2018a}, and either an AGN or composite according to the selection criteria by \citet{Hirschmann2019}.  The other C~IV emitters show a similar agreement with the typical AGN EWs and line ratios; in particular, RXCJ2248-ID is remarkably close to GHZ2 in all considered diagrams, similar to the case of the line-ratio diagnostics discussed above. 

Finally, we checked whether the non-detection of the very high-ionization lines \neiv\ and \nev\ discriminates between AGN and star-formation powered emission in the case of GHZ2. These lines have ionization potentials that are higher than those of any of the other UV emission features detected in GHZ2, and, in particular, \nev\ with a ionization energy of $\sim$ 97 eV is a robust marker of accretion onto a SMBH \citep[e.g.,][]{Schmidt1998,Gilli2010}.
We find that our EW limits are not stringent enough to rule out AGN emission in GHZ2. The limit on EW([Ne~V]) in GHZ2 is much higher than the typical values measured in both narrow-line ($\sim$8 \AA) and broad-line AGN ($\sim$1-3 \AA) at z$\sim$1 \citep{Mignoli2013}. Similarly, CIV-selected AGNs at $1.5 \leq z \leq 3$ have EW([Ne~IV])$\simeq$1-5 \AA\ \citep{Mignoli2019}, lower than the 3-$\sigma$ limit of 16 \AA\ we measure in GHZ2. On the one hand, GHZ2 shows properties similar to these obscured AGNs, that have typical EW(C~IV)$\simeq$30-60 \AA, and EW(C~III])$\simeq$15-30 \AA. On the other hand, the position of GHZ2 on the C~IV/He~II versus C~IV/C~III] plane falls just outside the region occupied by most of the aforementioned AGNs, which is well described by pure AGN models \citep[Fig. 5 in][]{Mignoli2019}.

These findings are consistent with the tests shown in Figs~\ref{fig_SFAGN1} and ~\ref{fig_SFAGN2}, and do not resolve the ambiguity on the nature of GHZ2. Instead, line-ratio diagnostics using very high-ionization lines lines can provide more stringent constraints. We explored several combinations, finding that the most useful diagram exploits the O~III]/He~II versus [Ne~IV]/N~IV] ratios which are compatible only with star formation according to the models by F16 and G16 (Fig.~\ref{fig_SFAGN_highion}). However, we cannot consider this test to be conclusive, since GHZ2 is still marginally compatible with the AGN models by NM22 and it has ratios consistent with those of GNz11 which shows AGN features \citep{Maiolino2023}. 
We also explored other diagnostics based on the [Ne~IV]/C~III] or [Ne~V]/C~III] ratios \citep{Scholtz2023}, and on Ne53=[Ne~V]/[Ne~III] \citep{Cleri2023}, finding that they are not conclusive because the available limits for GHZ2 are compatible with either very low-metallicity stellar populations or AGNs. In particular, GHZ2 has log(Ne53)$<$-0.31 at 3-$\sigma$, and log(O III]/H $\beta$)=0.72$\pm$0.11 \citep{Zavala2024,Calabro2024} compatible with the \textquote{composite} zone defined by \citet{Cleri2023}.
The \oiiibowen\ fluorescence line may point to GHZ2 being a Seyfert-like AGN. However, as already mentioned in Sec.~\ref{subsec:OIIIbowen}, the nondetection of the [Ne~V] line is puzzling in this respect.

To summarize, the available data do not allow us to ascertain whether the dominant ionization source within GHZ2 is star formation or AGN accretion. On the one hand, the high EWs of carbon and oxygen lines point to an AGN contribution.  On the other hand, the measurable line ratios, and in particular [Ne~IV]/N~IV], favor the star-formation case according to the F16 and G16 models, but are compatible with both scenarios according to the NM22 models.

\section{A star-formation scenario for GHZ2}\label{sec:SFSCENARIO}
The nondetection of very high-ionization lines and the similarity between GHZ2 and other C~IV emitters in terms of UV line ratios motivate us to explore the possible physical conditions of GHZ2 under the hypothesis that star-forming regions are the main source of ionizing photons.

\begin{table}
\renewcommand{\arraystretch}{1.2}
\caption{Physical properties of GHZ2$^a$}
\begin{center}
\begin{tabular}{| c | c |}
\hline
$M_{\rm UV}$ & $-20.53 \pm 0.01$  \\
\hline
UV slope &  $-2.46 \pm 0.08$  \\
\hline
log$(M_{\rm star}/{\rm M}_{\odot})$& 9.05$^{+0.10}_{-0.25}$     \\
\hline
SFR  (M$_{\odot}$~yr$^{-1}$) &  5.2$^{+1.1}_{-0.6}$ \\
\hline
sSFR (Gyr$^{-1}$)&  4.7$^{+5.1}_{-1.0}$   \\
\hline
$\Sigma_{\rm SFR}$ (M$_{\odot}$~yr$^{-1}$~kpc$^{-2}$) & 75 $\pm$ 4 \\
\hline
$\Sigma_{M}$ (M$_{\odot}$~pc$^{-2}$) & $16.2^{+1.1}_{-5.4} \times 10^3$ \\
\hline
A$_V$ (mag) & 0.04$^{+0.07}_{-0.03}$ \\
\hline
12 + log(O/H) & $7.26^{+0.27}_{-0.24}$ \\
\hline
log\,$U$ & $-1.78 \pm 0.28$  \\
\hline
\end{tabular}\label{proptable}
\end{center}
\small $^a$UV slope and magnitude measured from the continuum in the NIRSpec spectrum (Sect.~\ref{subsec:lya}). Metallicity and ioniziation parameter estimated from UV emission lines following \citet{Curti2023a}, and \citet{Mingozzi2022}, respectively  (Sect.~\ref{sec:met}). All other quantities obtained through SED-fitting using \textsc{BAGPIPES} (Sect.~\ref{sec:globalproperties}).
\end{table}

\subsection{Global properties from photometry}\label{sec:globalproperties}
We first derive global properties of GHZ2 by performing spectral energy distribution (SED) fitting of the observed photometry with \textsc{BAGPIPES} v. 1.0.3 \citep{Carnall2018,Carnall2019b}. The templates are based on BPASS v. 2.2.1 stellar models with an upper-mass cutoff of the IMF of 300~M$_{\odot}$ \citep{Stanway2018}, and nebular emission computed self-consistently with \textsc{CLOUDY} \citep{Ferland2013} as described by \citet{Carnall2018}. We assume a double power law star-formation history, and a \citet{Charlot2000} dust attenuation model with power-law index $0.3\leq n \leq 2.5$. 

After correcting for magnification, we find a star-formation rate (SFR) of $5.2^{+1.1}_{-0.6}$~M$_{\odot}$~yr$^{-1}$ and a stellar mass log\,$(M_{\rm star}/{\rm M}_{\odot}) = 9.05^{+0.10}_{-0.25}$, implying a specific star-formation rate sSFR = $4.7^{+5.1}_{-1.0}$~Gyr$^{-1}$. In agreement with the measured blue UV slope, GHZ2 is found to have very low dust extinction ($A_V = 0.04^{+0.07}_{-0.03}$~mag). While these values are obtained solely on the basis of photometry and we defer a detailed spectrophotometric fitting to future work, we note that the best-fit range for log\,$U$ ($-2.7$ to $-1.4$) and metallicity ($Z=0.02$--0.17~Z$_{\odot}$) are in reassuring agreement with the estimates based on emission lines that will be discussed below. We caution, however that broad-band observations of the optical rest-frame will be needed to obtain stronger constraints on the stellar mass which may be subject to systematics due to the large range of the M/L$_{UV}$ values \citep{Santini2023}. Somewhat reassuringly, as discussed in \citet{Zavala2024}, a fit performed jointly on photometry and MIRI spectroscopy yields consistent values for the stellar mass and SFR, and an estimate of the mass-weighted Age (=$28^{+10}_{-14}~$Myr) which is poorly constrained when using only the broad-band UV photometry. 

GHZ2 has an extremely small size, with an effective radius of $R_e = 105 \pm 9$~pc \citep[corrected for magnification,][]{Yang2022b}. An even smaller radius (=34$\pm 9$~pc, delensed) was measured by \citet{Ono2023}, who have shown that its morphology can also be fitted with a two-component model with a PSF-like plus compact ($\approx$42~pc, delensed) galaxy  profile.
By conservatively adopting the estimate from \citet{Yang2022b}, we obtain $\Sigma_{\rm SFR} =75 \pm 4$~M$_{\odot}$~yr$^{-1}$~kpc$^{-2}$ and $\Sigma_{M} = 16.2^{+1.1}_{-5.4} \times 10^3$~M$_{\odot}$~pc$^{-2}$, while $\sim$6-9 times larger values are obtained using the $R_e$ values estimated  by \citet{Ono2023} for the two- and single-component fits, respectively.

The properties of the source are summarized in Table.~\ref{proptable}.

\begin{figure*}
\centering
\includegraphics[trim={4.0cm 0.2cm 3.2cm 0.5cm},clip,width=18cm,keepaspectratio]{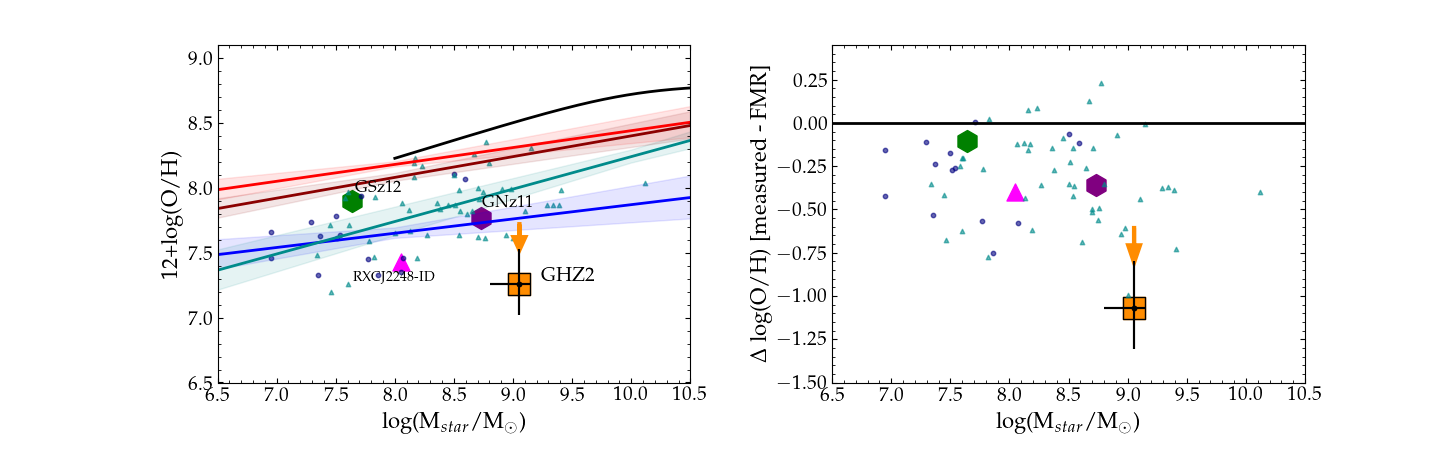}
\caption{\textit{Left:} the position of GHZ2 and other objects on the gas-metallicity versus stellar-mass plane (symbols as in Fig.~\ref{fig_SFAGN1}) compared to the mass-metallicity relations at $z=0$ \citep[black line;][]{Curti2020}, $z \approx 2$ (light red), and $z \approx 3$ (dark red) from \citet[][]{Li2023}, $z \approx 4$--9 \citep[cyan;][]{Nakajima2023} and $z\approx 6$--10 \citep[blue;][]{Curti2023}. The orange arrows indicate the upper limit on metallicity of 0.1\,Z$_{\odot}$ for GHZ2, which is the highest among available estimates. Blue and cyan points indicate single sources at $z > 6$ from \citet{Curti2023} and \citet{Nakajima2023}, respectively. The shaded regions enclose the scatter around the best-fit relations. \textit{Right:} deviation from the fundamental metallicity relation by \citet{Curti2023} as a function of stellar mass for GHZ2 and the other single sources with the same symbols as in the left panel.} 
\label{fig_MZR}
\end{figure*} 

\subsection{Metallicity and ionization properties}\label{sec:met}
Detailed modeling of the nebular conditions requires measuring line ratios of doublets sensitive to temperature or density \citep[e.g.,][]{Berg2019,Kewley2019} which are unresolved by the NIRSpec prism. Nonetheless, several scaling relations and criteria have been proposed in the literature to provide estimates of metallicity and ionizing conditions from UV emission lines when a thorough modeling is not possible. In fact, the strong \civ\ emission in GHZ2 is immediately suggestive of  low metallicity, as this line is only observed in star-forming galaxies with $Z\lesssim0.2$\,Z$_{\odot}$ \citep{Mingozzi2023}. We obtain a quantitative evaluation using the Ne3O2 = \neiii/\oii\ index which is considered a good proxy for oxygen abundance \citep[e.g.,][]{Maiolino2008}. We adopt the relation by \citet{Curti2023a} which extends to metallicity $< 0.1$\,\zzsun, obtaining  \logoh $= 7.26^{+0.27}_{-0.24}$, corresponding to $Z = 0.038^{+0.033}_{-0.016}$\,Z$_{\odot}$.  The more conservative Ne3O2 versus metallicity relation by \citet{Bian2018} yields a slightly larger value of \logoh $= 7.57^{+0.15}_{-0.10}$ ($Z=0.075^{+0.031}_{-0.015}$\,Z$_{\odot}$). Considering that the Ne3O2 index might be affected by collisional de-excitation of [OII] at high density, we explored other indicators. Reassuringly, the same metallicity range is obtained with the calibrations by \citet{Mingozzi2023} based on the C~III/O~III ratio ($\sim 0.05$\,\zzsun) and EW(C~III) ($\sim 0.1$\,\zzsun). 
The variance among different estimators is likely due to the intrinsic degeneracies among the abundance of oxygen and other elements and the physical conditions of the emitting gas. Nonetheless, we can safely assume a metallicity for GHZ2 in the range 3--10\% \zzsun.

When considering the metallicity obtained with Ne3O2 following \citet{Curti2023a}, we find a significant offset from the mass-metallicity relations (MZR) measured at different redshifts (Fig.~\ref{fig_MZR}, left panel), including the range $z \approx 6$--10 \citep[][]{Curti2023}. However, GHZ2 would still be  consistent with the MZR according to the more conservative estimate of 0.1\,Z$_{\odot}$. A more significant deviation is found compared to the fundamental metallicity relation \citep[FMR;][]{Mannucci2010}. As shown in the right panel of Fig.~\ref{fig_MZR}, GHZ2 is offset by $> 0.6$~dex from the FMR by \citet{Curti2023} considering all aforementioned metallicity estimates. In fact, given its mass and SFR, GHZ2 would be expected to have a metallicity of $\sim 0.5$\,Z$_{\odot}$ if following the FMR. 

Finally, we can infer the ionization parameter on the basis of the  relations between log\,$U$ and line properties measured by \citet{Mingozzi2023} on the CLASSY sample. Among the  equations proposed by \citet{Mingozzi2023}, we exploit the one based on the C~IV/C~III ratio, which is calibrated against the [O~III] $\lambda5007$/[O~II] $\lambda3727$ ratio that traces the intermediate/high-ionization zone. The C~IV/C~III ratio has the advantage that consecutive ionization states of the same element are less prone to potential biases owing to different abundances in GHZ2 and the CLASSY sources. We find a high ionization parameter, log\,$U = -1.78 \pm 0.28$, where the uncertainty includes the scatter in the fitting equation reported by \citet{Mingozzi2023}. The analysis of MIRI observations of GHZ2 yields to estimates for both the metallicity ($Z=0.02-0.17\,Z_\odot$) and the ionization parameter (log\,$U >$-2) consistent with our findings \citep{Zavala2024}.

A low metallicity and a high ionization parameter are typical of strong C~IV emitters. As a comparison, \rxc\ have a metallicity 4--8\% solar \citep{Topping2024}, with even lower oxygen abundances for \AStark\ \citep[2--4\%;][]{Stark2015b} and \JIzotov\ \citep[$\sim 2.5$\%;][]{Izotov2024}. Similarly, their ionization parameters are extreme, ranging from log\,$U = -1.7$ (\JIzotov) to $\sim -1$ (\AStark\ and \rxc). Less extreme, but still comparable values have been measured in GNz11 \citep[0.08--0.12\,\zzsun, log\,$U \approx -2$;][]{Bunker2023}. Instead, the carbon-enhanced source GSz12 appears to be both more metal enriched \citep[$\sim 0.16$\,\zzsun;][]{DEugenio2023} and with a significantly lower ionization parameter (log\,$U < -2$ from the C~IV/C~III ratio, considering a 3$\sigma$ upper limit on the C~IV line).

\subsection{Is GHZ2 carbon- or nitrogen-enhanced?}

We used \pyneb\ \citep{Luridiana2012, Luridiana2015} to estimate the C/O and the N/O abundance ratios, with the task \verb|getIonAbundance|. 
For electron density we consider the discrete values [$10^3$, $5 \times 10^3$, $10^4$, $5 \times 10^4$, and $10^5$]~cm$^{-3}$, and for the electron temperature we consider 1.5, 2, 2.5, 3 $\times 10^4$~K. We note that a high electron density log(n$_e$/cm$^{-3}$)$>$3 is favoured by the combined analysis of the MIRI data and of the ALMA observations of GHZ2  from \citet{Bakx2022} \citep{Zavala2024}. For C/O, we assume it can be inferred from the C$^{2+}$/O$^{2+}$ ratio to which we apply an ionization correction factor (ICF). Using the calibration by \cite{Berg2019}, which depends on the ionization parameter, and for a metallicity 0.05\,Z$_{\odot}$, we find an ICF of 1.2.   We use the line ratio \ciii/\oiii\  and the grid of electron densities and temperatures to estimate the total C/O. For each assumed temperature we performed 1000 Monte Carlo simulations perturbing the observed fluxes by their corresponding uncertainties. 
We obtain values ranging from log(C/O) $= -0.94$ (for $n_e = 10^5$~cm$^{-3}$ and $T = 1.5 \times 10^4$~K) to log(C/O) $= -0.53$ (for $n_e = 10^3$~cm$^{-3}$ and $T = 3.0 \times 10^4$~K). These estimates imply that GHZ2 has a subsolar carbon abundance of [C/O] $\approx -0.68$ to $-0.27$~dex \citep[assuming log(C/O)$_{\odot} = -0.26$ from][]{Asplund2009}.

We explore the N/O abundance ratio adopting our 7$\sigma$ fiducial detection of the \niii~line and assuming that the oxygen abundance is dominated by the O$^{2+}$ state and the total N/O is derived as N/O $\approx$ (N$^{2+}$ + N$^{3+}$)/O$^{2+}$. We use the line ratios \niv/\oiii\ and N~III] $\lambda 1750$/\oiii, and the same grid of electron densities and temperatures as above. 
In all considered cases the derived N/O
exceeds the solar value, ranging from log(N/O) $ = -0.29$ (for $n_e =10^5$~cm$^{-3}$, $T = 3.0 \times 10^4$~K) to log(N/O) $= -0.2$ (for $n_e=10^3$~cm$^{-3}$, $T=1.5 \times 10^4$~K); these are a factor of $\sim 4$--5  higher than the solar abundance \citep[log(N/O)$_{\odot} = -0.86$;][]{Asplund2009}. However, the measured nitrogen abundance relies on the detection of the \niii\ line which, as discussed in Sec.~\ref{sec:spectrum}, needs further assessment considering the low SNR and the dependence on the estimated continuum level. In fact, when considering the conservative 3$\sigma$ upper limit discussed above in place of a detection for \niii, we get an upper limit of log(N/O) $\lesssim -0.4$.   

The above-mentioned estimates indicate a remarkable similarity between GHZ2 and the recently discovered class of nitrogen-enhanced objects. In fact, GNz11 and \rxc\ have both a supersolar nitrogen abundance and a subsolar C/O \citep{Senchyna2023,Topping2024}, and similar cases of high C/N ratios are also discussed by \citet{Isobe2023b}. 

\subsection{Our proposed scenario: highly ionizing radiation sources in a dense environment}
Assuming a star-forming origin of the ionizing radiation inside GHZ2, the elemental abundance described above hints to conditions that are significantly different from the typical cases of low- and intermediate-redshift galaxies. In fact, available models struggle at reproducing the high EW of UV lines observed in GHZ2 and in other sources \citep[e.g.,][]{Nanayakkara2019,Saxena2020,Olivier2022}.  Additional sources of ionizing photons may include X-ray binaries \citep{Schaerer2019,Garofali2024}, very massive stars \citep{Lecroq2024,Upadhyaya2024}, soft X-rays emitted by hot gas within young stellar clusters \citep{Oskinova2022}, and shock-ionized gas \citep{Jaskot2016}. 

In this context, the \oiiibowen\ fluorescence line detected in GHZ2 may provide insights into the highly ionizing sources responsible for the He~II emission and the extreme EW of the carbon lines. While detailed modeling of this uncommon emission feature is beyond the scope of the present work, it must be stressed that it indicates the presence of dense, highly ionized gas within the object \citep[e.g.,][]{Deguchi1985}. Its emission mechanism requires a large amount of He-ionizing photons ($> 54.4$~eV) recombining into the unobservable He~II Ly$\alpha$ line at 303.782~\AA, which is not surprising in our case considering the detection of the He~II Balmer-$\alpha$ line at 1640~\AA\ with EW $\approx 5$~\AA.

The C and N abundances provide a signature of early enrichment dominated by burning of hydrogen through the CNO cycle \citep[e.g.,][]{Isobe2023b}. In particular, the nitrogen enrichment can be explained in the same context as the \textquote{excess} of ionizing photons by the presence of supermassive stars in a densely clustered environment \citep{Charbonnel2023,Senchyna2023,Vink2023,Marques-Chaves2024}.

The abundance patterns in GHZ2 and other similar sources are compatible with them being formation sites of the stellar populations that will end up in globular clusters and other very dense environments studied in the local universe 
showing similar nitrogen excess \citep[e.g.,][]{DAntona2023}. In this respect, it is notable that GHZ2 has both a high $\Sigma_{\rm SFR}$ (log($\Sigma_{\rm SFR}$~[M$_{\odot}$~yr$^{-1}$~kpc$^{-2}$]) $\approx 1.9$) and a high
$\Sigma_{M}$ (log($\Sigma_{M}$~[M$_{\odot}$~pc$^{-2}$] ) $\gtrsim 4$), making this object an ideal host of gravitationally bound stellar clusters. Similar dense conditions have been observed through gravitational lensing at $z \approx 2$--8, e.g., the Sunburst \citep{Vanzella2022a}, Sunrise \citep{Vanzella2023}, and Firefly Sparkle \citep{Mowla2024}, along with the recently discovered $z \approx 10.2$ Cosmic Gems \citep[][]{Adamo2024} which shows extremely dense stellar clusters located within a 50~pc physical region. In this scenario, the luminosity and density of GHZ2 
can be explained by the presence of young massive stellar clusters at the peak of their formation, 
the majority of them confined within a barely resolved 100~pc scale \citep[e.g.,][]{Adamo2024}. Following \citet{Langeroodi2023}, the compact size of GHZ2 can also explain its significant deviation from the FMR in a scenario where centrally concentrated star formation is ignited by infall of pristine gas \citep[see also][]{Tacchella2023}.

While \rxc\ is more extreme in terms of SFR and stellar mass density than GHZ2, the similarities with GHZ2 are also striking. The two objects have very similar line ratios and EWs, and are both dense and highly ionized objects with comparable metallicity and abundance patterns. These findings suggest that GHZ2 belongs to a population of very high-redshift objects that are undergoing an intense phase of  star formation in a dense and compact volume. It is tempting to speculate that the remnants of this phase will end up in dense stellar environments at low redshift. 

Its characteristics, combined with the blue UV slope $\beta = -2.39$ measured from the spectrum (which suggests little dust opacity), are those expected in scenarios in which the objects populating the bright end of the UV luminosity function have undergone a \textquote{blue monster} phase that effectively pushed away the previously formed dust through radiation driven winds \citep{Ferrara2023, Fiore2023}.

Objects in the same evolutionary phase as GHZ2 may also significantly contribute to cosmic reionization. The analysis of \citet{izotov2023} shows a relation between low metallicity, high N/O, low C/O, and significant leakage of Lyman continuum (LyC) photons. In addition, \citet{Schaerer2022a} and \citet{Kramarenko2024} found a correlation between a high C~IV/C~III ratio and a nonzero escape fraction of ionizing photons: GHZ2 has C~IV/C~III $\approx 3$, higher than the threshold of 0.75 characterizing strong LyC leakers according to the aforementioned analysis. Indication of a significant escape fraction of ionizing photons has also been found at $z\approx 3$ in C~IV emitters by \citet{Mascia2023}, with these objects \citep[and strong-line emitters in general, e.g.,][]{Castellano2023b} also having a high photon production efficiency. In this respect, deeper and higher resolution spectra of GHZ2 will be able to investigate its escape fraction through the analysis of the MgII $\lambda$2800 doublet \citep[e.g.,][]{Chisholm2020}, which is currently undetected to a limit (EW$\lesssim$10\AA) that prevents this kind of constraints. 

\section{Conclusions}\label{sec:summary}

We report here the spectroscopic confirmation at $z=12.34$ of GHZ2/GLASS-z12. The spectroscopic redshift is in remarkable agreement with the estimates obtained from NIRCam photometry \citep[e.g.,][]{Castellano2022b,Naidu2022b,Harikane2022b}, lending support to the accuracy of {\it JWST}-based photometric selections of high-redshift galaxies, at least as far as bright objects are concerned. Together with similar results obtained from NIRSpec follow-up spectroscopy \citep[e.g.,][]{ArrabalHaro2023b}, this provides a crucial confirmation that the relatively large density of bright galaxies at $z \gtrsim 9$ is real and deserves detailed investigation in order to understand the earliest phases of galaxy and structure formation.

The spectrum of GHZ2 shows strong N~IV, C~IV, He~II, O~III, C~III, O~II, and Ne~III emission lines. The prominent C~IV line puts GHZ2 in the category of strong C~IV emitters \citep[][]{Stark2015b,Izotov2024}. In fact, GHZ2 is the most distant, brightest, and most massive member of this recently discovered class of objects. Assessing the main source of ionizing photons from UV spectroscopy is known to be challenging, in particular for high-redshift objects likely dominated by young, low-metallicity stellar populations. In fact, our extended comparisons with theoretical models spanning a large range in metallicity and ionization parameter are somewhat inconclusive. We find that, similarly to other C~IV emitters and other very-high-redshift objects, the large EWs of UV lines found in GHZ2 seem to imply an AGN-like ionizing spectrum, while the UV line ratios are broadly compatible both with AGN emission and with star formation in a very low-metallicity environment. 

While it is not possible to draw a firm conclusion from the aforementioned tests, we consider the star-forming case slightly favoured in light of the low [Ne~IV]/N~IV] ratio, and of the comparison with C~IV emitters which are considered to be most likely star-forming galaxies. In fact, the line ratios of GHZ2 are remarkably similar to those of \rxc, whose spectrum shows no AGN signatures \citep{Topping2024}, and of  \AStark\ and \JIzotov\ that are well reproduced by highly ionizing stellar populations \citep{Stark2015b,Izotov2024}.

Regardless of the nature of the dominant ionizing flux, we found that GHZ2 has a very low metallicity (below 10\% solar) and a high ionization parameter (log\,$U > -2$). The N/O abundance is found to be 4--5 times the solar value, while the C/O is subsolar, similar to a number of recently discovered high-redshift objects \citep{Topping2024,Isobe2023b}. Given its small effective radius ($R_e \approx 100$~pc), GHZ2 has a high $\Sigma_{\rm SFR}$ and a high stellar mass density similar to gravitationally bound stellar clusters; it is intriguing to speculate that GHZ2 is undergoing a phase of intense star formation in a dense configuration that may evolve into the nitrogen-enhanced stellar populations of globular clusters and other dense environments that are observed at low redshifts.

The origin of the copious amounts of ionizing photons in objects such as GHZ2 is currently unknown, but scenarios of dense star formation at very low metallicity including supermassive stars and high-mass X-ray binaries have the potential to also explain the atypical abundance patterns and the high-ionization spectra. A detailed investigation of the rare detection of the \oiiibowen\ fluorescence line in GHZ2 can provide further insight into its sources of ionizing photons and their local environment. The high C~IV/C~III ratio and the abundance patterns of GHZ2 are also suggestive of a high escape fraction of ionizing photons. A comprehensive search for sources in this evolutionary stage can reveal if they play a significant role in the reionization of the inter-galactic medium. 

We caution, however, that in-depth studies will be needed to consolidate the proposed scenario. The N/O ratio needs to be assessed on the basis of a robust detection of the N~III line in a higher-resolution spectrum considering the strong dependence of our measurement on the extrapolated continuum level at its position. Similarly, a high-resolution spectrum is needed to estimate the density and temperature of the ionized gas, to ascertain the presence of broad components due to AGNs or stellar winds, and to separate nebular and stellar contributions to the C~IV emission.

The remarkable brightness of GHZ2 makes it accessible to a wealth of follow-up strategies, as showcased by the results described in this paper and by the MIRI detection discussed in a companion paper \citep{Zavala2024}.  As such, GHZ2 has the potential to become a reference object for understanding galaxy formation at only 360~Myr after the Big Bang.

\begin{acknowledgments}
We  thank  the  referee  for  the  constructive comments that helped us improve the manuscript. We thank K. Nakajima and A. Feltre for kindly providing updated tables of their line-emission models, and A. Carnall for the useful feedback on BAGPIPES SED fitting. We thank E. Giallongo and E. Piconcelli for the interesting discussions. We thank Tony Roman (Program Coordinator) and Glenn Wahlgren (NIRSpec reviewer) for the assistance in the preparation of GO-3073 observations.  
This work is based on observations made with the NASA/ESA/CSA {\it James Webb Space Telescope (JWST)}. The JWST data presented in this article were obtained from the Mikulski Archive for Space Telescopes (MAST) at the Space Telescope Science Institute. The specific observations analyzed are associated with program JWST-GO-3073 and can be accessed via \dataset[DOI]{https://doi.org/10.17909/4r6b-bx96}. We acknowledge financial support from NASA through grant JWST-ERS-1342. Support was also provided by the PRIN 2022 MUR project 2022CB3PJ3 – First Light And Galaxy aSsembly (FLAGS) funded by the European Union – Next Generation EU, and by INAF Mini-grant ``Reionization and Fundamental Cosmology with High-Redshift Galaxies." A.V.F. is grateful for financial assistance from the Christopher R. Redlich Fund and numberous other donors. C.G. and P.R. acknowledge support through grant MIUR2020 SKSTHZ. CM acknowledges support by the VILLUM FONDEN under grant 37459 and the Carlsberg Foundation under grant CF22-1322. The Cosmic Dawn Center (DAWN) is funded by the Danish National Research Foundation under grant DNRF140. K.G. and T.N. acknowledge support from Australian Research Council Laureate Fellowship FL180100060.

\end{acknowledgments}

%

\facilities{JWST, VLT:Kueyen}


\software{A-PHOT \citep[][]{Merlin2019}, Astropy \citep{Astropy2013,Astropy2018,Astropy2022}, BAGPIPES \citep{Carnall2018,Carnall2019b}, JWST Calibration Pipeline version 1.13.4 \citep{JWSTPipeline}, Matplotlib \citep{Hunter2007}, \pyneb~\citep{Luridiana2012, Luridiana2015}, Specutils (\url{https://specutils.readthedocs.io/en/stable/}),  T-PHOT \citep[][]{Merlin2015,Merlin2016a}}



\appendix

\section{VLT X-SHOOTER spectroscopy}\label{appendix:XSHOT}
GHZ2 was observed by X-SHOOTER under program 110.244H.001 (PI E. Vanzella). The observations were performed in nodding mode with $11 \times 0.9''$ slits and 900~s per single exposure in the near-infrared arm. The total observing time was 37~hrs, corresponding to an on-target exposure time of 30~hrs. 

Data reduction was performed as by \citet{Castellano2022a} with the official XSHOOTER pipeline \citep{Modigliani2010}, v.~3.3.5, using the associated raw calibrations from the ESO archive. Sky subtraction was done using the XSHOOTER nodding strategy, combining frames acquired at two different positions in the sky for each OB execution. A fixed boxcar window was used to extract the target on the rectified 2D spectra, to ensure that the target was correctly localized along the slit (\texttt{extract\-method=LOCALIZATION} and \texttt{localize\-method=MANUAL} in pipeline recipe \texttt{xsh\_scired\_slit\_nod}).

The 1D extracted spectra were corrected to the barycentric reference frame in vacuum and combined with the data-analysis package \textsc{Astrocook} \citep{Cupani2020}. The combined spectra were rebinned to a resolution of 4.8~\AA\ per pixel. 

\begin{figure*}
\centering
\includegraphics[trim={0.1cm 0.1cm 0.1cm 0.1cm},clip,width=18cm,keepaspectratio]{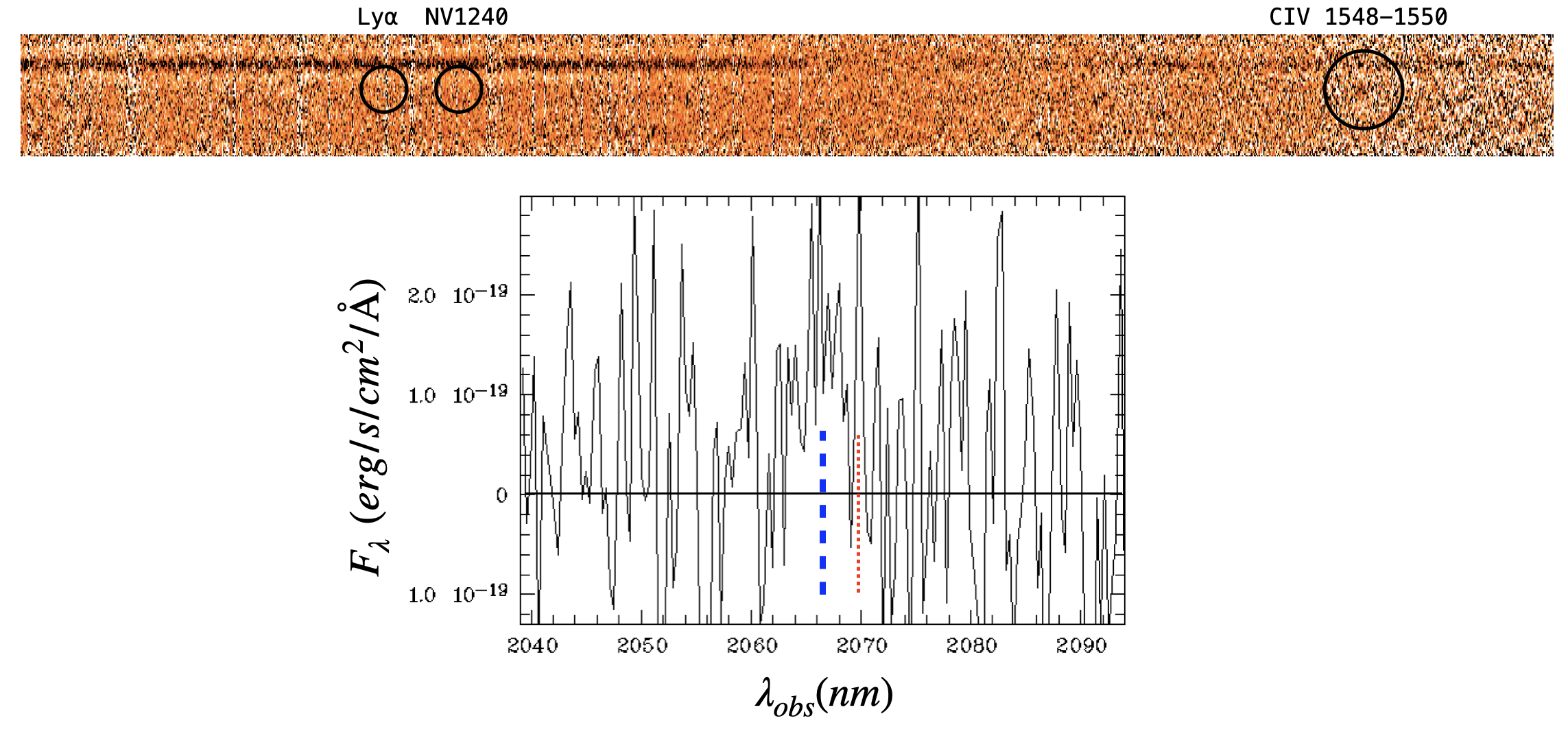}
\caption{\textit{Top:} the 2D X-SHOOTER spectrum of GHZ2 with highlighted the expected positions of Ly$\alpha$ and N~V $\lambda 1240$, and the position of the detected C~IV $\lambda 1548$ line. \textit{Bottom:} the extracted 1D spectrum at the position of the C~IV $\lambda 1548$ emission (blue dashed line). The red dotted line indicates the expected position of the C~IV $\lambda 1551$ component.}
\label{fig_CIVXSHOOTER}
\end{figure*}

We detect at SNR = 4.4 an emission feature at $\lambda = 20665.16$~\AA, which is compatible with one of the two components of the \civ\ emission detected with NIRSpec (Fig.~\ref{fig_CIVXSHOOTER}). We interpret it as the C~IV $\lambda 1548$ line which is the brightest component of the doublet, implying $z=12.348$, consistent with the NIRSpec redshift.  The measured flux is $(0.78 \pm 0.18) \times 10^{-18}$~erg~s$^{-1}$~cm$^{-2}$, and we estimate a FWHM=390 $\pm$ 160~km~s$^{-1}$ rest-frame. This estimate is lower than observed in broad-line AGNs \citep[e.g.,][]{Shen2011} and it is consistent with a stellar wind origin of the line \citep[e.g.,][]{Chisholm2019}. However, the low SNR of the detection clearly prevents any firm assessment of the line profile which will require higher resolution JWST spectroscopy.

The SNR implies a 3$\sigma$ limit of $0.6 \times 10^{-18}$~erg~s$^{-1}$~cm$^{-2}$ for the second component, which leaves the ratio between the two poorly constrained. Following \citet{Lemaux2009}, we expect slit losses in the X-SHOOTER spectrum to be small ($\sim 10$--15\%) on the basis of the slit dimension,  the compact size of the source ($< 0.1''$), and the median seeing ($\sim 0.8''$). Considering that the total \civ\ flux detected in the (aperture-corrected) NIRSpec spectrum is $(2.57 \pm 0.6) \times 10^{-18}$~\ergs, the flux of the C~IV $\lambda 1548$ line implies a relatively high ratio between the two components of the doublet \citep[as in, e.g., \AStark;][]{Stark2015b} and/or the presence of an additional broad component \citep[e.g.,][]{Berg2019,Llerena2021} which is not detectable in the X-SHOOTER spectrum and shall be investigated with NIRSpec high-resolution spectroscopy.

No other features are found at the expected position of other emission lines, consistent with  the fluxes measured in the NIRSpec spectrum and the noise level in the X-SHOOTER one. Interestingly, the higher resolution of the X-SHOOTER spectrum enables measurements close to the Lyman break, which is severely affected by damping wing absorption at the PRISM resolution yielding to a 3$\sigma$ upper limit on N~V $\lambda 1240$ of $0.6 \times 10^{-18}$~erg~s$^{-1}$~cm$^{-2}$.

\bibliography{biblio3}{}
\bibliographystyle{aasjournal}



\end{document}